\documentclass{article}

\usepackage{amsmath}
\usepackage{comment}
\usepackage{subcaption}
\usepackage{graphicx}
\usepackage{makecell}
\usepackage{url}
\usepackage{xcolor}

\graphicspath{{Images/}}

\begin{document}

\title{A General Language for Modeling \\ Social Media Account Behavior}

\author{Alexander C. Nwala\thanks{Corresponding author. Current affiliation: William \& Mary. Email: acnwala@wm.edu} \and Alessandro Flammini \and Filippo Menczer\\Observatory on Social Media\\Indiana University, Bloomington}

\sloppy
\maketitle

\begin{abstract}
Malicious actors exploit social media to inflate stock prices, sway elections, spread misinformation, and sow discord. To these ends, they employ tactics that include the use of inauthentic accounts and campaigns. 
Methods to detect these abuses currently rely on features specifically designed to target suspicious behaviors.
However, the effectiveness of these methods decays as malicious behaviors evolve.
To address this challenge, we propose a general language for modeling social media account behavior. Words in this language, called BLOC, consist of symbols drawn from distinct alphabets representing user actions and content.
The language is highly flexible and can be applied to model a broad spectrum of legitimate and suspicious online behaviors without extensive fine-tuning. 
Using BLOC to represent the behaviors of Twitter accounts, we achieve performance comparable to or better than state-of-the-art methods in the detection of social bots and coordinated inauthentic behavior.
\end{abstract}

\section{Introduction}

The widespread use of social media makes them a prime target for exploitation by bad actors.
Efforts to inflate the popularity of political candidates~\cite{ratkiewicz2011detecting} with social bots~\cite{ferrara2016rise}, influence public opinion through the spread of disinformation and conspiracy theories~\cite{Lazer-fake-news-2018, grinberg2019fake}, and manipulate stock prices through coordinated campaigns~\cite{cresci2019cashtag,pacheco2020uncovering} have been widely reported.
The threats posed by malicious actors are far-reaching, endangering democracy~\cite{schiffrin2017disinformation, woolley2018computational}, public health~\cite{tasnim2020impact, allington2020health, covaxxy-misinfo}, and the economy~\cite{fisher2013syrian}.
In response, researchers have developed various tools to detect malicious inauthentic accounts.

However, we are in an arms race. With new detection methods and prevention mechanisms from platforms, malicious actors continue to evolve their behaviors to evade detection.
For example, consider the evolution of social bots: in the early days, spam bots were easy to identify because they often lacked meaningful profile information and/or demonstrated naive behaviors~\cite{yardi2010detecting,lee2011seven}.
In recent years, bot accounts have become more sophisticated. Some display detailed profiles, either stolen from other users or generated by deep neural networks~\cite{nightingale2022ai}.
Some mimic human actions and build social connections~\cite{cresci2020decade}. 
Others adopt strategies such as \textit{coordinated inauthentic behaviors}.\footnote{\url{about.fb.com/news/2018/12/inside-feed-coordinated-inauthentic-behavior}}
Such coordinated behaviors appear to be normal when inspected individually, but are centrally controlled to achieve some goal~\cite{pacheco2020uncovering}.

The arm race has spawned a series of more complex detection methods~\cite{cresci2020decade,mazza2019rtbust,pacheco2020uncovering}.
An important limitation of these methods is that they rely on features crafted specifically to target previously observed malicious behaviors~\cite{sayyadiharikandeh2020detection}.
These features may not generalize well to other suspicious behaviors.
For example, methods designed to detect sophisticated social bots tend to overlook coordinated behaviors, and vice versa~\cite{yang2019arming}.
Existing methods also become less useful when facing novel malicious actors, unless the features are adjusted accordingly.

\begin{figure}
  \centering
  \includegraphics[width=\textwidth]{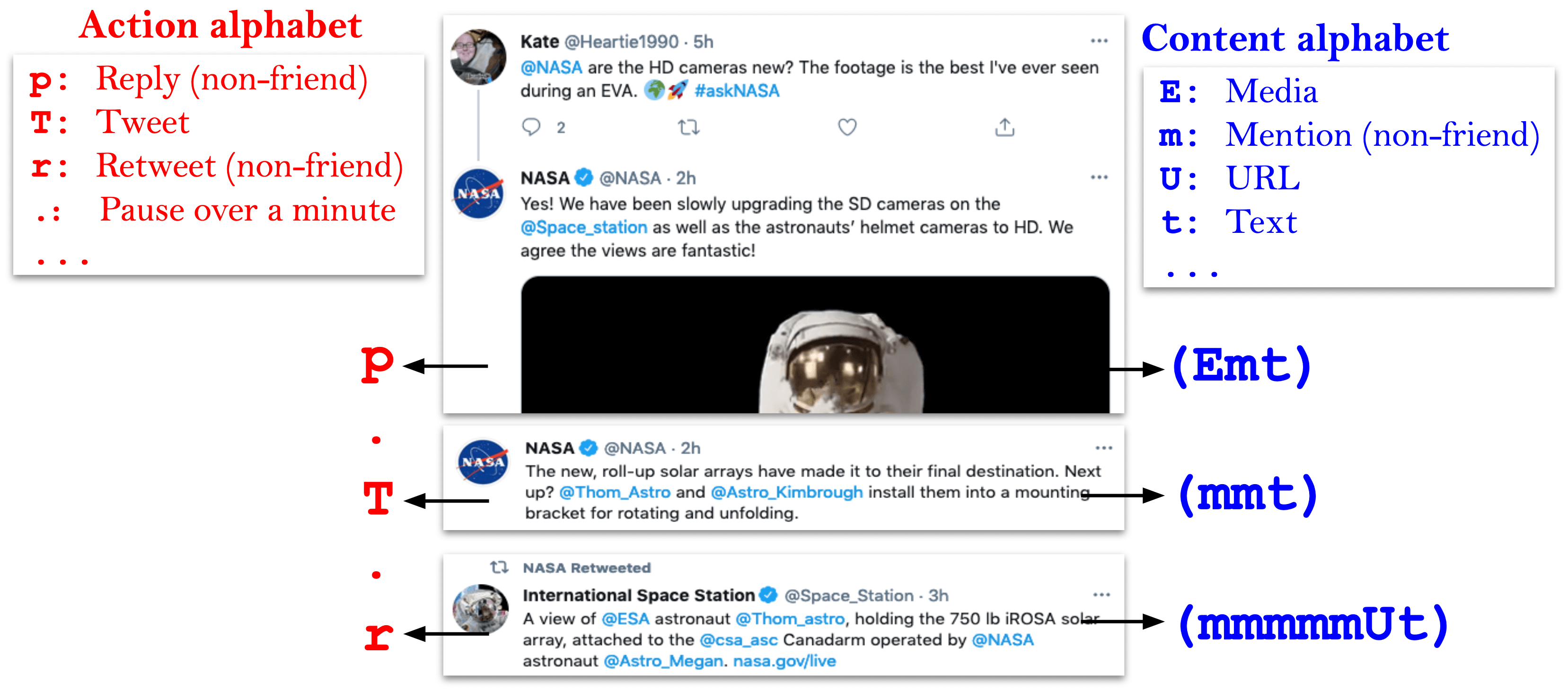}
  \caption{BLOC strings for a sequence of three tweets (a reply, an original tweet, and a retweet) by the \texttt{@NASA} account.
  Using the \textit{action} alphabet, the sequence can be represented by three word $p.T.r$ separated by dots. Using the \textit{content} alphabet, it can be represented by these three words $(Emt)(mmt)(mmmmmUt)$ enclosed in parentheses. See Sec.~\ref{sec:bloc_main} for details.} 
  \label{fig:nasa_bloc}
\end{figure}

To address this challenge, we propose the Behavioral Language for Online Classification (BLOC), a general language that represents social media account behaviors.
BLOC words consist of symbols drawn from distinct alphabets representing an account's actions and content. 
As an example, Fig.~\ref{fig:nasa_bloc} illustrates possible representations of a sequence of tweets by the official Twitter handle for NASA. 
The BLOC language is highly flexible in that it can represent a broad spectrum of legitimate and suspicious behaviors without extensive fine-tuning.
In this paper we show that meaningful behavioral patterns emerge from such representations, facilitating tasks related to the classification of social media accounts.

To demonstrate the effectiveness of BLOC, we evaluate it on social bot and coordinated behavior detection tasks, together with previous methods specifically designed for each of the two tasks. To the best of our knowledge, BLOC is the only existing representation that has been applied to both tasks. Although methods based on BLOC use significantly fewer features than state-of-the-art methods ---making them much more efficient--- they yield better or comparable performance.

\section{Related work}

We can think of at least two dimensions to characterize inauthentic online behaviors: automation and coordination. Accounts could be automated but independent, or coordinated but closely managed by humans, or both automated and coordinated, and everything in between.  Below we outline research aimed to detect inauthentic behaviors along these dimensions.
Note that not all automated or coordinated behavior is necessarily inauthentic or malicious. For example, some self-declared bots are harmless or even useful; and some  grassroots campaigns may use coordination to promote beneficial social movements.

\subsection{Automation}
\label{subsec:related_work_automation}

The behavioral spectrum of social media account automation has human behavior at one end and bot-like behavior at the opposite end.
Somewhere in between are ``cyborgs''~\cite{chu2012detecting, chu2010tweeting}, accounts that cycle between human and bot-like behaviors.

Various machine-learning methods have been proposed for identifying specific kinds of automated behavior. 
These methods typically utilize some combination of features such as social network structure, content/profile characteristics, and temporal patterns~\cite{ferrara2016rise}.

Some researchers have focused on characterizing human behavior online.
Wood-Doughty et al. studied one million accounts to explore how different demographic groups used Twitter~\cite{wood2017does}.
This was based on the assumption that user behavior is reflected by indicators such as profile personalization, temporal information, location sharing, user interaction, and devices.
He et al. provided a method for identifying five classes of behaviors on Twitter: individual, newsworthy information dissemination, advertising and promotion, automatic/robotic, and other activities~\cite{he2014identifying}.
Researchers have also studied human behavior across other social media platforms.
Maia et al. represented YouTube users as feature vectors over a vocabulary consisting of number of uploads, videos viewed, channels visited, system join date, age, and so on~\cite{maia2008identifying}.
They then clustered the users into predefined profiles such as small community member, content producer, and content consumer.
Benevenuto et al. studied the online behavior of over 37 thousand users who accessed four social networks (Orkut, MySpace, Hi5, and LinkedIn) by analyzing their clickstream data~\cite{benevenuto2009characterizing}. 

\label{subsec:related_work_detect_bot_like_behavior}

On the other end of the automation spectrum are social bots~\cite{ferrara2016rise}. 
A common theme of the literature is to build algorithms to distinguish bot-like and human accounts~\cite{cresci2020decade}, which requires representing the account characteristics first.
The rich information obtained from social media platforms makes it possible to describe accounts along many different dimensions. 
Depending on the types of the target accounts, existing methods use profile information~\cite{yang2020scalable}, content~\cite{yardi2010detecting,cresci2017social,cresci2017exploiting}, actions~\cite{mazza2019rtbust}, social network~\cite{beskow2018bot}, and temporal signatures~\cite{mazza2019rtbust}. 

Another common approach is to combine account characteristics from different dimensions in the same model~\cite{lee2011seven,davis2016botornot,varol2017online,gilani2017bots,yang2019arming,sayyadiharikandeh2020detection}.
Botometer,\footnote{\url{botometer.org}} for example, is a publicly available supervised machine learning system that extracts over 1,000 features from a Twitter account's profile, content, sentiment, social network, and temporal activity.

Digital DNA (DDNA), proposed by Cresci et al.~\cite{cresci2017social,cresci2017exploiting}, is the most similar method to BLOC.
DDNA encodes each account as a pair of strings of symbols representing actions and content, respectively. It then considers accounts with long common substrings as bots.
While BLOC similarly uses sequences of symbols to encode actions and types of content, it differs from DDNA in a major way, namely by capturing pauses.
BLOC can segment strings into words that represent sequences of actions or content symbols separated by pauses. Words may capture distinct behavioral patterns, while the absence of long pauses could be revealing of automated behaviors. Accounts can thus be represented as word vectors, allowing for similarity measures beyond string matching.
Another difference is that DDNA truncates repetition; for example, a single $U$ character represents one or more URLs, whereas BLOC can capture repetitions (e.g., $UUU$) to emphasize different behaviors. This can help detect repetitive behaviors, such as long sequences of retweets, typical of certain inauthentic accounts.

\subsection{Coordination}

Malicious social bots evolve in sophistication over time, making them more effective and harder to detect.
In some cases, it is not sufficient to study individual accounts. A group of inauthentic accounts can be coordinated by a single entity, whether their behavior are human-controlled or automated. 
These kinds of sophisticated deception can only be detected through observations at the group level~\cite{cresci2020decade}.
This has led to multiple research efforts to detect malicious coordinated behaviors.

While individual bot detection aims to separate individual human and bot-like accounts, coordination detection involves clustering suspiciously similar accounts into groups~\cite{pacheco2020uncovering}.
Appropriate definitions of similarity measures are subjective and vary across different studies.
A common choice is to focus on the temporal dimension, with the action time series of different accounts compared directly~\cite{chavoshi2016debot,keller2017manipulate} or modeled using temporal point processes~\cite{sharma2021identifying}. 
Other similarity measures focus on duplicated or partially matched text~\cite{assenmacher2020two,vargas2020detection} or on shared retweets~\cite{nizzoli2021coordinated}.
Some methods focus on specific components of the content, such as embedded links, hashtags, and media~\cite{pacheco2020uncovering,keller2020political,giglietto2020coordinated,giglietto2020takes,vargas2020detection}.
Account profile information can also be used to identify similar accounts~\cite{fazil2020socialbots}.
Finally, it is possible to aggregate similarity measures based on different criteria~\cite{magelinski2021synchronized}.

These methods typically extract account features designed to target specific suspicious behavioral patterns~\cite{pacheco2020uncovering}. 
BLOC encodes behavioral information into features that can be used to calculate similarities without a predefined target behavior. 
As a result, BLOC is versatile and can be applied to characterize a broad spectrum of behaviors.
We next provide an in-depth introduction to BLOC.

\section{Behavioral Language for Online Classification}
\label{sec:bloc_main}

The central component of BLOC is a collection of two alphabets: \textit{actions} and \textit{content}. Each consists of a set of symbols that represent activities or traits. Collectively, these alphabets encode behaviors that can be utilized to build models for various tasks.

BLOC has several language parameters, shown in Table~\ref{tab:bloc_parameters_states}. Different combinations of values for these parameters correspond to different languages and representations. Below we discuss these parameters in detail, noting  recommended values based on extensive experiments. In Sections~\ref{sec:behavior_characterize} and \ref{sec:eval} we apply different BLOC representations to various tasks. 

\begin{table}
\centering
\caption{BLOC language parameters.}
\begin{tabular}{llll}
\hline
\textbf{Param.} & \textbf{Context} & \textbf{Explanation} & \textbf{Values} \\ \hline
$p_1$ & Pauses    & Session delimiter threshold             &  Time        \\ 
$p_2$ & Pauses    & Time granularity              &  $f_1(\Delta)$ or $f_2(\Delta)$  \\
$p_3$ & Word      & Use sessions for content words    &  Yes or No                    \\ 
$p_4$ & Word      & Tokenization                  &  N-gram or pause                 \\ 
$p_5$ & Word      & Sort symbols                  &  Yes or No                    \\ 
$p_6$ & Word      & Word truncation length &  Integer                    \\ \hline
\end{tabular}
\label{tab:bloc_parameters_states}
\end{table}

\subsection{BLOC alphabets}

Let us illustrate how to generate BLOC strings drawn from the alphabets, for an arbitrary Twitter account \texttt{@Alice}. Note, however, that BLOC is platform-agnostic.

\subsubsection{Action alphabet}

The \textit{action} alphabet includes two sets of \textit{action} and \textit{pause} symbols. An action symbol characterizes a single post by an account with a symbol as outlined below:
\begin{itemize}
\setlength{\itemsep}{0pt}
\setlength{\parskip}{0pt}
    \item [$T$:] Post message
    \item [$P$:] Reply to friend
    \item [$p$:] Reply to non-friend
    \item [$\pi$:] Reply to own post 
    \item [$R$:] Reshare friend's post
    \item [$r$:] Reshare non-friend's post
    \item [$\rho$:] Reshare own post
\end{itemize}
For example, the string $Tp\pi R$ indicates that \texttt{@Alice} posted a tweet, then replied to a non-friend, followed by a reply to herself, and finally retweeted a friend.

The \textit{pause} symbols characterize the pauses between consecutive actions. 
Pauses provide additional context for actions. For example, actions taken with very short (e.g., less than a second) or highly regular pauses could indicate automation~\cite{ghosh2011entropy}.

Let us first define $\Delta$ as the time between two consecutive actions. 
Based on parameter $p_2$, we have two possible pause alphabets defined by functions that map $\Delta$ values to symbols. The function $f_1$ is defined as: 
\begin{equation}
  f_1(\Delta) = 
    \begin{cases}
      \text{no symbol}   & \text{if $\Delta < p_1$} \\
      . & \text{otherwise} \\
    \end{cases}
  \label{eqn:bloc_time}
\end{equation}
where $p_1$ is a \textit{session} delimiter threshold.
A session is thus defined as a maximal sequence of consecutive actions separated by pauses shorter than $p_1$. 
Sessions are important because they provide natural word boundaries for tokenizing BLOC words (see Section \ref{subsec:bloc_vector_models}).
We recommend using a value of a minute or less for $p_1$ in Eq.~\ref{eqn:bloc_time}. 

As an illustration, let us punctuate \texttt{@Alice}'s string of actions ($Tp\pi R$) with pause symbols using $f_1$ and $p_1 = 1$ minute. Say that Alice pauses 2.5 minutes between the first tweet and the reply to a non-friend, then 50 seconds pass until her self-reply, and finally she waits 3 days before the final friend retweet. The resulting BLOC string would be $T.p\pi.R$, indicating three sessions whose boundaries are marked by the dots. 

An alternative pause alphabet assigns different symbols to long pauses for better granularity. We discretize time into a logarithmic scale to represent a wide range of pauses, e.g., hours vs. days vs. weeks, by defining $f_2$ as:
\begin{equation}
f_2(\Delta) = 
  \begin{cases}
  \text{no symbol} & \text{if $\Delta < p_1$} \\
  t_h & \text{if $p_1 \leq \Delta < 1$ hour} \\
  t_d & \text{if 1 hour $\leq \Delta < 1$ day} \\
  t_w & \text{if 1 day $\leq \Delta < 1$ week} \\
  t_m & \text{if 1 week $\leq \Delta < 1$ month} \\
  t_y & \text{if 1 month $\leq \Delta < 1$ year} \\
  t_z & \text{otherwise.} \\
  \end{cases}
\label{eqn:bloc_time_long_pause}
\end{equation}
Using the same example as above, \texttt{@Alice}'s string of actions using the $f_2$ pause symbols with $p_1=1$ minute would be $T t_h p \pi t_w R$.

\subsubsection{Content alphabets}

The \textit{content} alphabet provides a lexical characterization of a post --- whether it contains text, links, hashtags, and so on. 
Unlike the \textit{action} alphabet, a single social media post can contain multiple \textit{content} symbols from the following list:
\begin{itemize}
    \setlength{\itemsep}{0pt}
    \setlength{\parskip}{0pt}
    \item [$t$:] Text
    \item [$H$:] Hashtag
    \item [$M$:] Mention of friend
    \item [$m$:] Mention of non-friend
    \item [$q$:] Quote of other's post
    \item [$\phi$:] Quote of own post
    \item [$E$:] Media object (e.g., image/video)
    \item [$U$:] link (URL)
\end{itemize}
As an illustration, let us imagine that \texttt{@Alice}'s first tweet only contains text; her reply to a non-friend has two images and one hashtag; her self-reply mentions one friend and has one link; and finally she retweets a post that mentions a non-friend. The resulting content string depends on the $p_3$ parameter. If sessions are not used, each action corresponds to a separate content word: $(t)(EEH)(UM)(m)$. Here the contents of the reply to a non-friend ($EEH$) and of the self-reply ($UM$) are separated, even though they were part of the same session. Using sessions, we get $(t)(EEHUM)(m)$. Note that parentheses separate content words, and the order of content symbols within a word is arbitrary and defined in the implementation.

\subsection{BLOC models}

BLOC strings can be used to build mathematical models for tasks such as online behavior characterization, bot detection, and coordination detection. 
Possible model classes include Markov chains and vector spaces.

\subsubsection{Language models}

A simple way to model an account with BLOC is a Markov chain. Each state in the model represents a BLOC symbol, with a transition link from state $s_t$ to state $s_{t+1}$ quantifying the probability $P(s_{t+1}|s_t)$ that symbol $s_{t+1}$ follows symbol $s_t$. For each account, we could perform multiple stochastic operations associated with Markov Chains. For example, we could predict the next action (e.g., tweet, share) of an account by computing $\arg \max_{s_{t+1}} P(s_{t+1}|s_{t})$ or estimate the likelihood that a sequence of actions were generated by an account, $P(s_1 \dots s_n) = \prod_{t=1}^{n-1} P(s_{t+1}|s_{t})$.

A more generalized language model, for example using deep-learning techniques~\cite{jurafskyspeech}, could be trained to generate sequences of BLOC symbols.

\subsubsection{Vector models}
\label{subsec:bloc_vector_models}

One could apply a deep-learning method, such as word2vec~\cite{mikolov2013distributed}, to embed BLOC strings into vector representations. However, the abstract vector space would fail to benefit from the interpretability of BLOC symbols.
Alternatively, we can obtain a vector representation by first tokenizing BLOC strings into words and then using these words directly as vector space dimensions.

Tokenization can be done using one of two methods, \textit{n-gram} or \textit{pause}, based on parameter $p_4$ (Table~\ref{tab:bloc_parameters_states}). 
The $n$-gram method generates tokens of fixed size $n$ by sliding an $n$-sized window over the BLOC string. Using $n = 2$, we generate bi-grams resulting in a vocabulary of two-symbol words. For example, given the action string $Tp\pi.r$ and the BLOC content string $(t)(EH)(U)(mm)$ with $n = 2$, we obtain the set of words $\{Tp$, $p\pi$, $\pi.$, $.r$, $tE$, $EH$, $HU$, $Um$, $mm\}$.

The pause method uses pauses to break BLOC action strings into words of variable length. In addition to serving as word boundary markers, pause symbols are included in the vocabulary as single-symbol words.
For content strings, individual posts mark word boundaries: all symbols in the same post form a single word. 
The symbols within each word may be sorted alphabetically depending on parameter $p_5$. 
To illustrate pause tokenization without sorting, given the same BLOC action string $Tp\pi.r$ and BLOC content string $(t)(EH)(U)(mm)$, we obtain the set of words $\{Tp\pi$, $.$, $r$, $t$, $EH$, $U$, $mm\}$. 

\begin{figure*}
  \centering
  \includegraphics[width=0.95\linewidth]{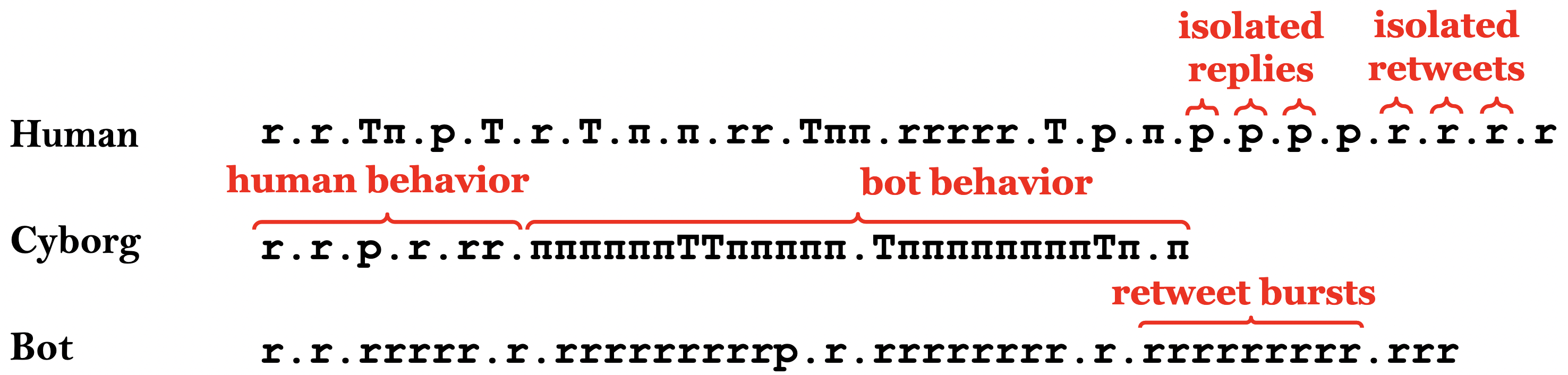}
  \caption{Illustrations of BLOC action strings ($p_1 = 1$ minute) for a human, a cyborg, and a bot Twitter account illustrating some behavioral differences across these individuals. If strings are tokenized using pauses, the human account has the shortest words (average length 1.35 vs. 3.88 for the cyborg and 4.0 for the bot) and is dominated by isolated retweets and replies. The cyborg account --- which we created to post threads of news updates --- exhibits both human (isolated posts) and bot behavior (thread bursts). The bot account mainly generates retweet bursts.}
  \label{fig:bloc_action_for_3}
\end{figure*}

Pause tokenization often results in long words, for example, the 13-symbol word $\pi\pi\pi\pi\pi\pi TT\pi\pi\pi\pi\pi$ from the \textit{cyborg} account in Fig.~\ref{fig:bloc_action_for_3}. Long words occur when the pauses between multiple consecutive actions are shorter than $p_1$, meaning that actions are performed in bursts, which often indicates automation. 
The distinction between, for example, $rrrr$ and $rrrrr$ is often not important, so instead of representing both as separate words in our vocabulary, we could truncate long words after a limit. For example, setting $p_6 = 4$ would truncate characters that repeat four or more times. The words $rrrr$, $rrrrr$, and $rrrrrr$ would all be replaced by $rrr+$.

After tokenization, we can represent any account as a vector of BLOC words. 
In a vector model, each account is represented as a point $(w_1, w_2,..,w_k)$ in a $k$-dimensional vector space where each dimension $i$ corresponds to a word. We wish to define a weight $w_i$ that represents how well an account is described by $i$. The number of times $f_i$ that word $i$ occurs in the BLOC representation of the account, known as term frequency (TF), is not very discriminative because some words, such as $t$ (text), may be common across all accounts. Therefore the term frequency is  multiplied by a second factor, called inverse document frequency (IDF), that captures how rare a word is across accounts. We use the TF-IDF weight~\cite{TFIDF} for account $a$ defined as follows:
\begin{equation}
  w_i(a) = f_i(a) \left(1 + \log \frac{D}{d_i} \right)
\end{equation}
where $d_i$ is the number of accounts with word $i$ and $D$ is the total number of accounts.
Finally, the vectors can be used to build bot or coordination detection systems. 

\section{Discriminative power of BLOC}
\label{sec:behavior_characterize}

BLOC lets us study behaviors at different levels of granularity. We may study different classes of accounts, such as humans vs.~bots. 
Or we might study different types of individual accounts within a class, for instance political vs.~academic human accounts or spam bots vs.~self-declared bots. 
In this section we demonstrate such a multi-resolution approach by characterizing the behavior of individual accounts and groups of accounts, both when their class labels are known and unknown.

\begin{figure}
    \centering
    \includegraphics[width=0.48\textwidth]{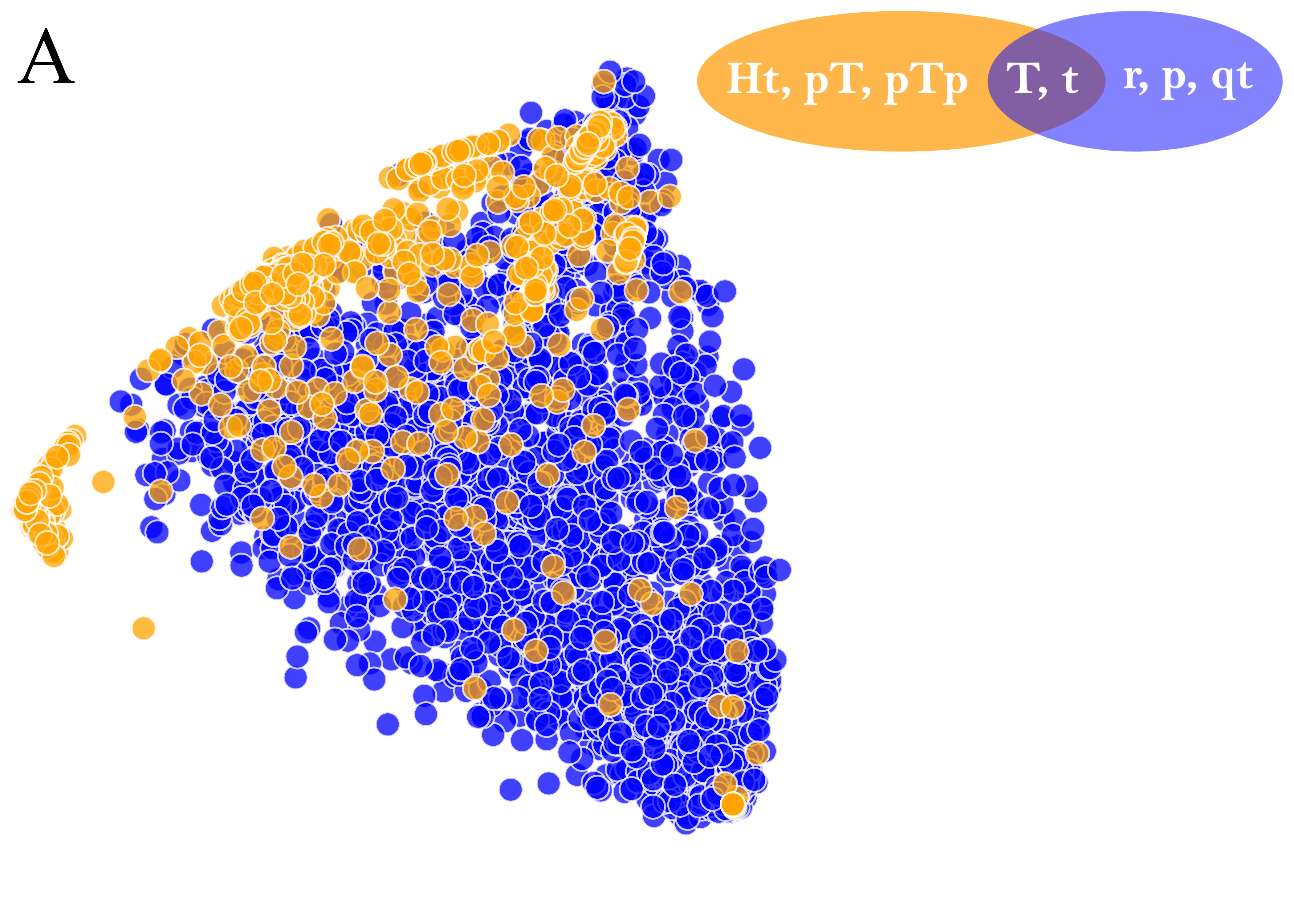}
    \includegraphics[width=0.48\textwidth]{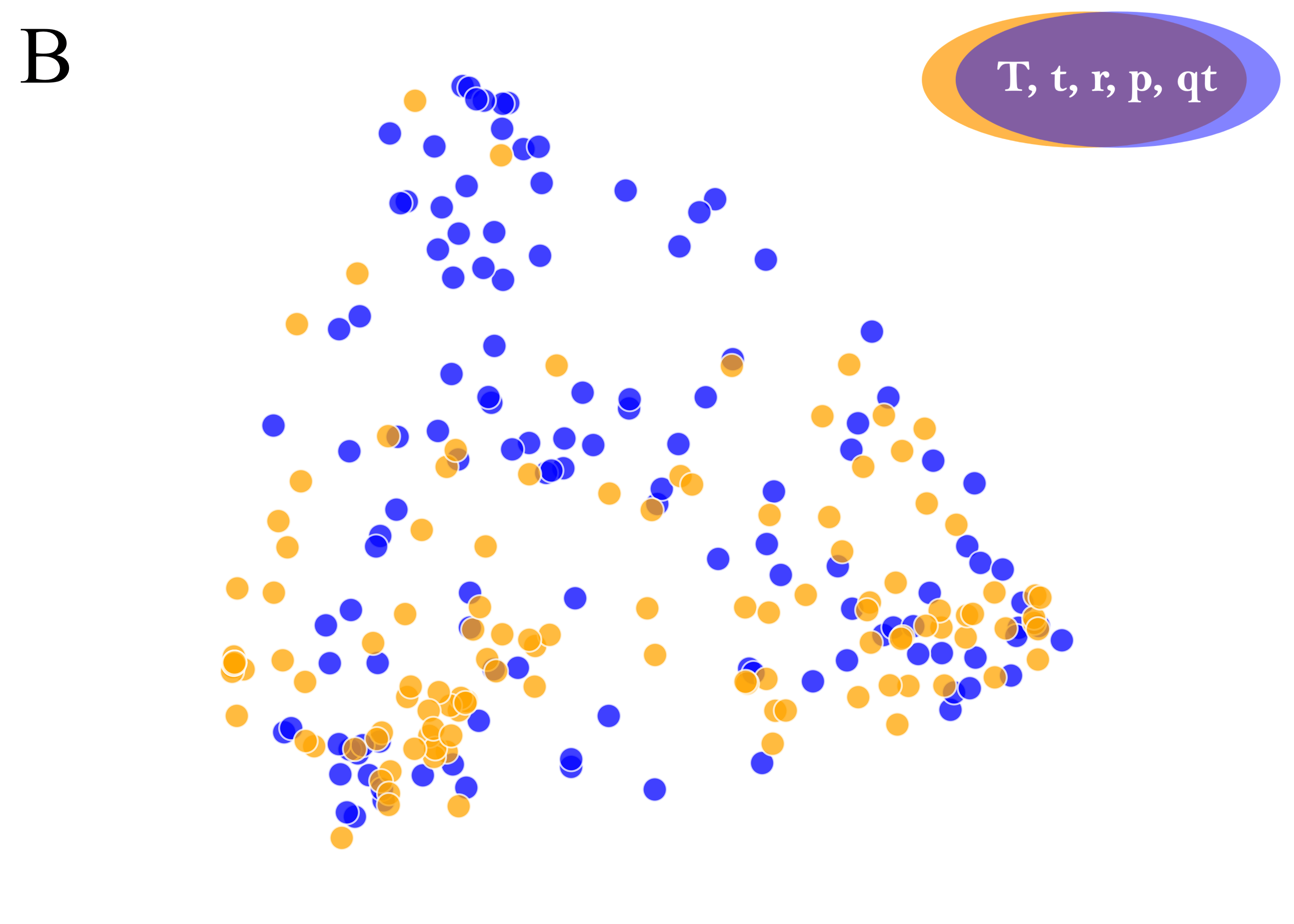}
    
    \includegraphics[width=0.48\textwidth]{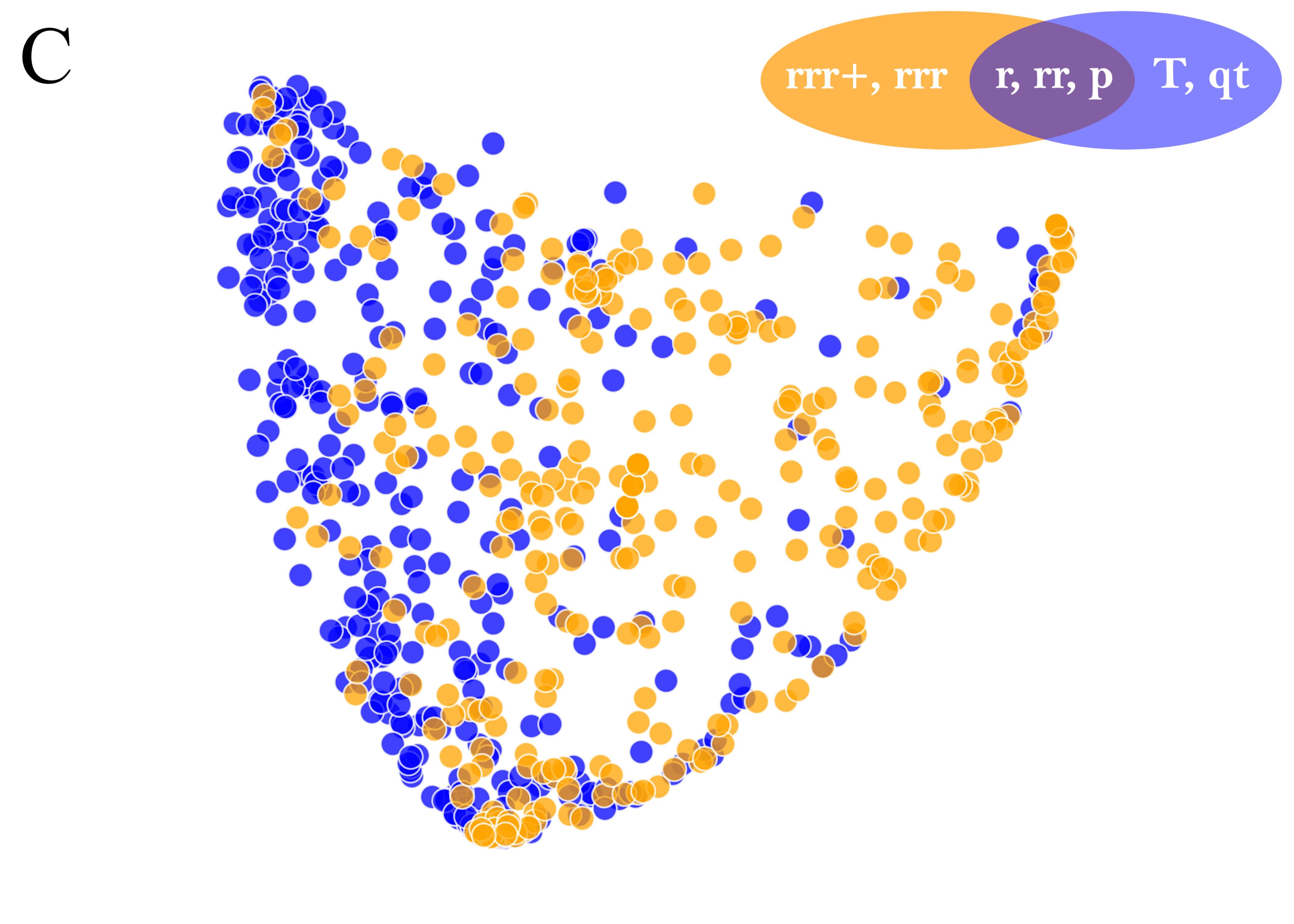}
    \includegraphics[width=0.48\textwidth]{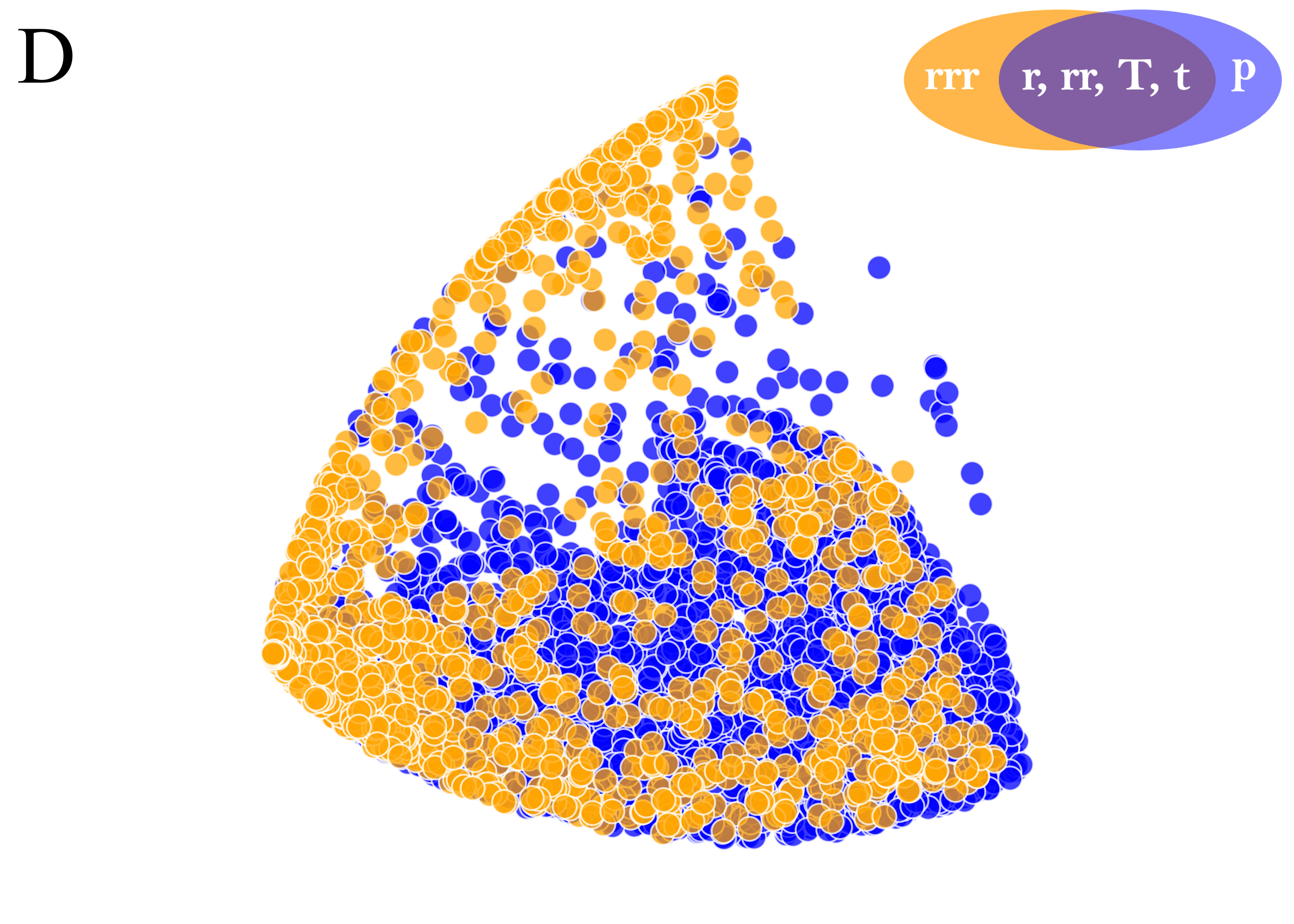}
    \includegraphics[width=0.48\textwidth]{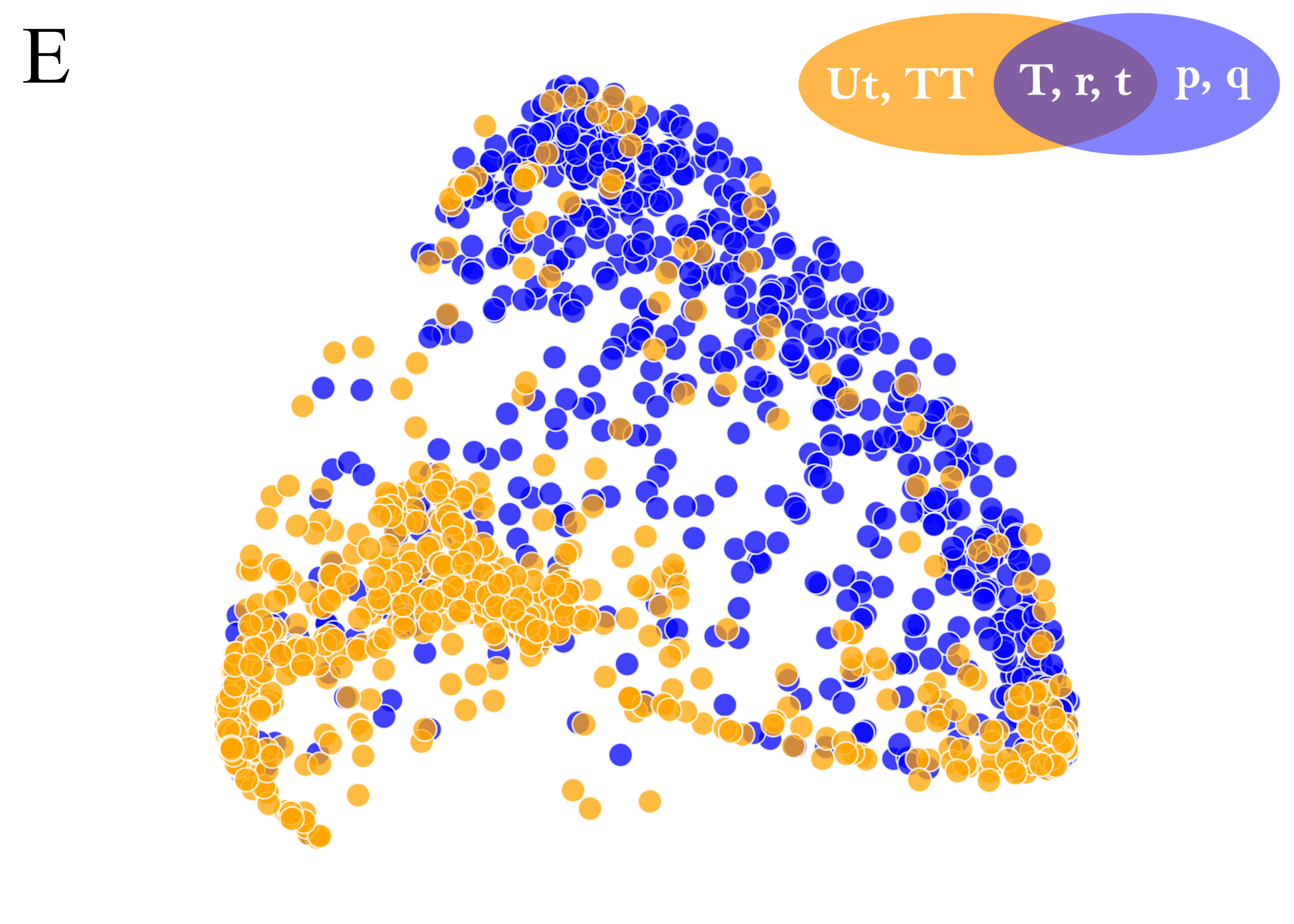}
    \includegraphics[width=0.48\textwidth]{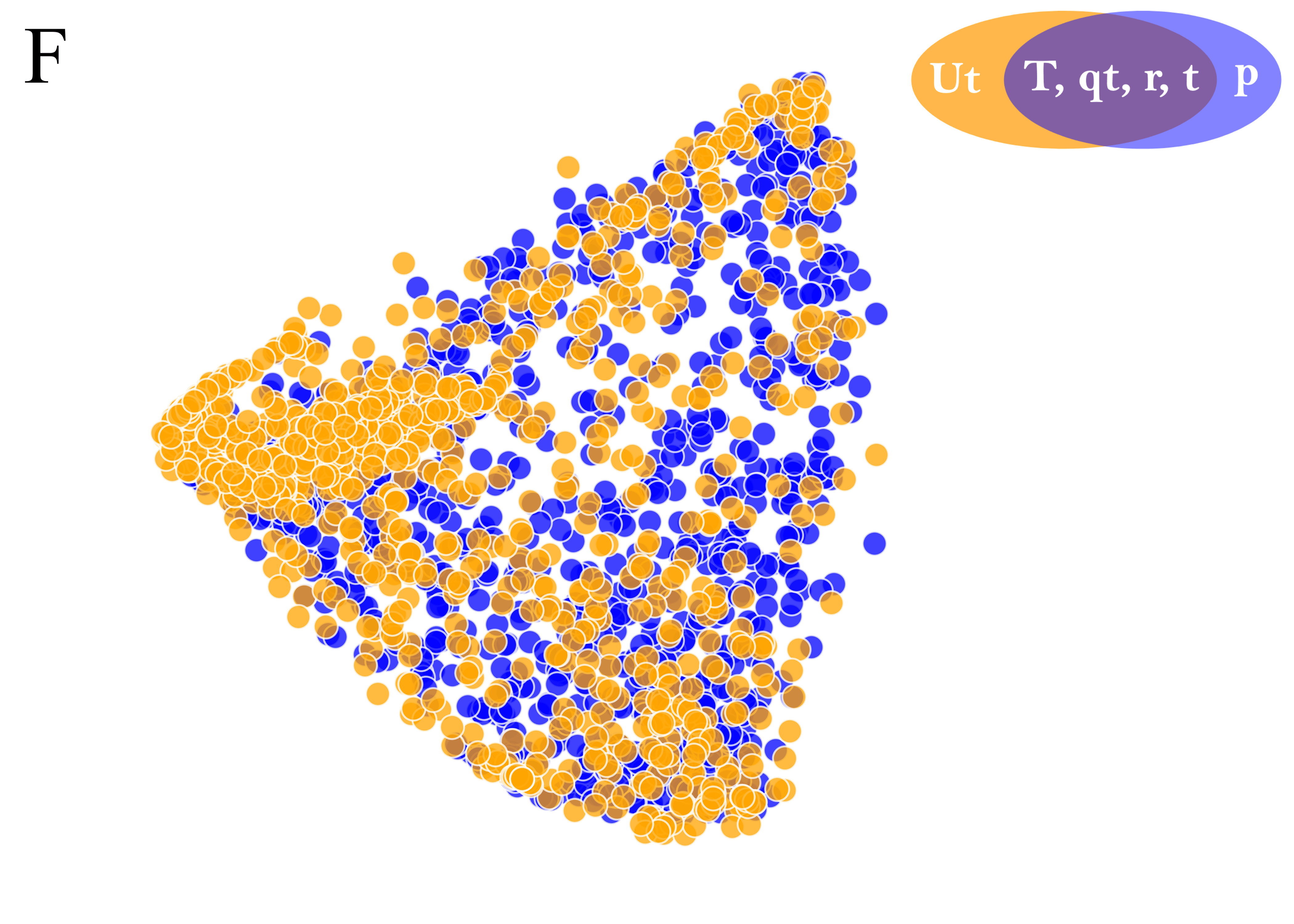}
    \caption{Two-dimensional PCA projections of BLOC TF-IDF vectors of accounts from six datasets that include both humans and bots (see Table~\ref{tab:bot_eval_dataset}): (\textbf{A})~cresci-17, (\textbf{B})~botometer-feedback-19,  (\textbf{C})~cresci-rtbust-19,  (\textbf{D})~cresci-stock-18,  (\textbf{E})~varol-17, and (\textbf{F})~gilani-17.
    From each of these datasets, we select an equal number of bot (orange) and human (blue) accounts. We use all the accounts in the minority class, and sample an equal number of accounts from the majority class. The Venn diagrams show the top five pause-delimited BLOC words for the bot and human accounts shown.}.
    \label{fig:bloc_bot_multi_src_pca}
\end{figure}

\subsection{Characterizing individuals and groups}

Fig.~\ref{fig:bloc_action_for_3} illustrates the behavioral differences between three individual accounts: a human account belonging to a journalist; a cyborg account used by one of the authors to post news updates, either manually or using a software script; and a spambot account identified by Mazza et al.~\cite{mazza2019rtbust}. 
These accounts are represented by their respective BLOC action  strings. 
We observe multiple differences. First, when we tokenize the strings into words separated by pauses, the human account has the shortest words, with mostly one-symbol words (e.g., $r$, $T$, $p$). This captures the fact that humans tend to rest between posts.
Second, the cyborg account exhibits a human substring with shorter words, followed by a bot substring created in a burst. 
Third, the bot account tends to amplify content with retweet bursts (e.g., $rrrrrrrrr$) rather than creating new content.

Let us shift our focus to studying groups of accounts. Fig.~\ref{fig:bloc_bot_multi_src_pca} presents a Principal Component Analysis (PCA) of the BLOC TF-IDF vectors of equal numbers of bot and human accounts from six different datasets (see Table~\ref{tab:bot_eval_dataset}). We observe that the bot and human accounts in the left column of the figure express more distinct behavioral patterns than those in the right column. Consequently, accounts in the left column have fewer words in common and are easier to separate. 
For example, while both bot and human accounts in Fig.~\ref{fig:bloc_bot_multi_src_pca}A tweet text-only ($t$) content, the bot accounts more often include hashtags ($Ht$). In Fig.~\ref{fig:bloc_bot_multi_src_pca}C, bots amplify content with burst of retweets ($rrr$, $rrr+$) unlike humans who create original content ($T$). In Fig.~\ref{fig:bloc_bot_multi_src_pca}E, bots share more external links ($U$) while humans tend to engage in conversations and commentary ($p$, $q$). 

In Fig.~\ref{fig:bloc_bot_multi_src_pca}B, bots and humans express similar behavioral traits: both classes have the same five top words. 
In Fig.~\ref{fig:bloc_bot_multi_src_pca}D and F, bots and humans share four of their five top words. The bot accounts are more likely to amplify content ($rrr$) and link to external websites ($Ut$) in Figs.~\ref{fig:bloc_bot_multi_src_pca}D and F, respectively, while their corresponding human accounts are more likely to engage in conversations ($p$). 
In summary, the figure suggests that the behaviors displayed by the humans tend to be consistent across datasets, whereas the bots have distinct behaviors based on the purpose for which they have been created. These findings are consistent with prior analysis based on ad-hoc features~\cite{sayyadiharikandeh2020detection}.
The BLOC representation is sufficiently powerful to capture significant differences between these behaviors.

\subsection{Behavioral clusters}
\label{subsec:id_coord_detect_vaccine}

When behavioral class labels are unavailable, we can characterize online behaviors in an unsupervised way, using BLOC to cluster accounts according to behavioral similarity.

We analyzed tweets collected between January 4 and September 30, 2021 from the CoVaxxy project,\footnote{\url{osome.iu.edu/tools/covaxxy}} which studies how online misinformation impacts COVID-19 vaccine uptake~\cite{covaxxy-misinfo}. The dataset~\cite{DeVerna_Pierri_2021} consists of over 200 million English-language tweets about COVID-19 and vaccines, posted by over 17 million accounts. The tweets were collected with 76 keywords and hashtags covering a variety of neutral (e.g., \textit{covid}), pro-vaccine (e.g., \textit{getvaccinated}), anti-vaccine (e.g.,  \textit{mybodymychoice}), and conspiratorial (e.g., \textit{greatreset}) topics.

Given the large number of accounts present in the dataset and the quadratic cost of pairwise comparison, we focused on the one thousand most active accounts each month. We based our definition of activity on the number of days in which an account posted tweets; to break ties (especially for accounts active every day), we used the total number of tweets an account posted during the collection period.

We applied a three-step, network-based method to identify clusters of accounts with highly similar behaviors. First, we generated BLOC TF-IDF vectors for each account using pauses to tokenize words, without sorting symbols, and truncating words ($p_6 = 4$). Second, we computed the cosine similarities among the  1,000 vectors. We built a network by linking only nodes (accounts) with similarity of at least 0.98 and removing singletons. This threshold ensures a focus on accounts with a suspiciously high level of similarity. Third, we applied the Louvain method to identify communities~\cite{Blondel}. 

\begin{figure*}
  \centering
  \includegraphics[width=0.98\linewidth]{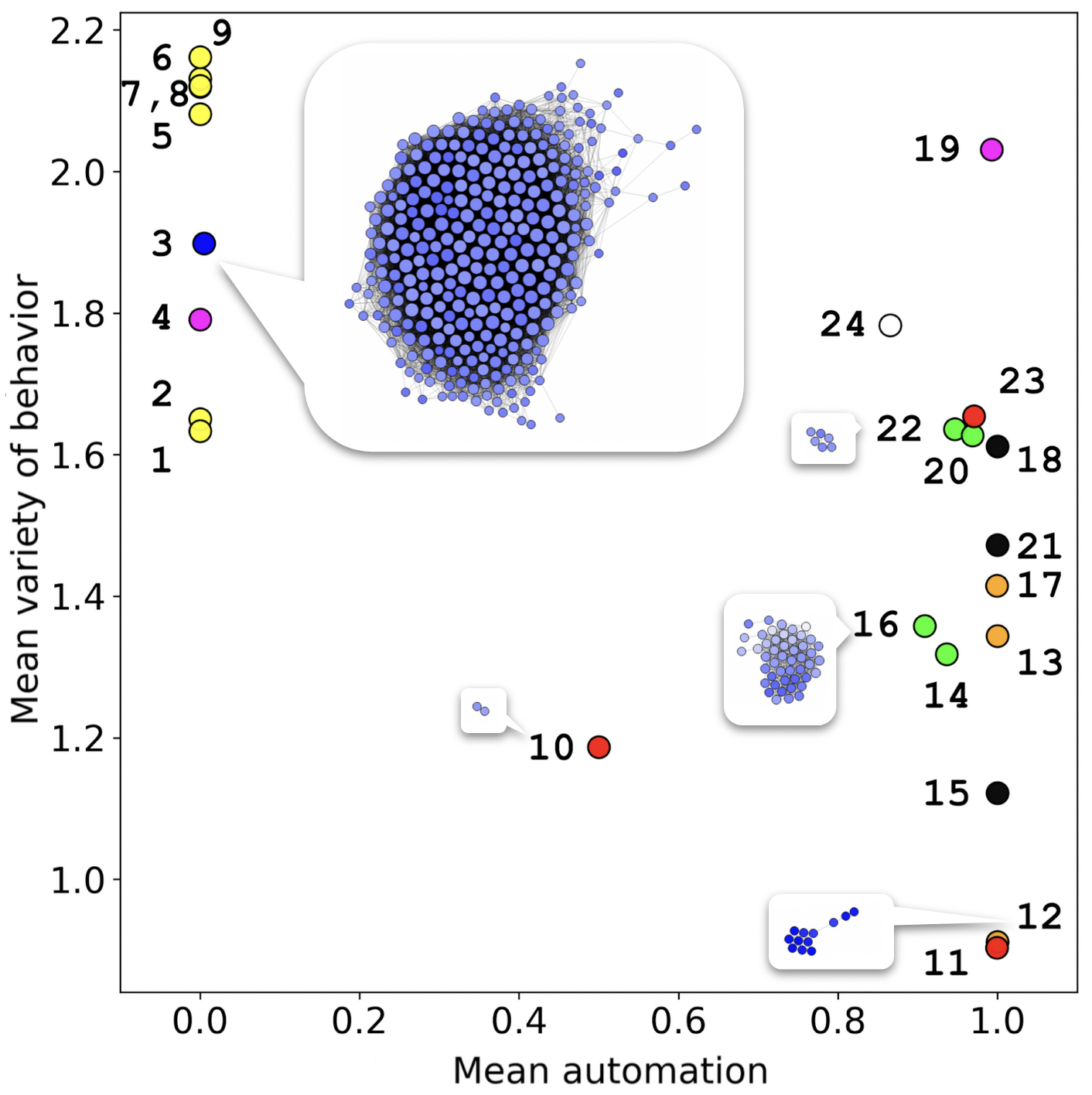}
  \caption{Mean variety of behavior vs. mean automation (see text) for 24 communities of accounts with highly similar behaviors. Each community is represented by a dot, colored according to manual classification (see text). A few selected communities are highlighted by visualizing the corresponding subnetworks, with node size and darker color representing degree and tweet count, respectively.}
  \label{fig:bloc_coord_composite}
\end{figure*}

This procedure was applied every month (January -- September) to produce nine behavioral similarity networks consisting of clusters of accounts with highly similar behaviors. Fig.~\ref{fig:bloc_coord_composite} visualizes 24 of the 163 identified clusters. In the figure, a single dot represents a cluster positioned on axes representing its \textit{mean variety of behavior} and \textit{mean automation} score. For a single account, we measured its variety of behavior by the entropy of its BLOC string (before tokenization). We estimated account automation by the fraction of times the account posted using the Twitter API. A user has to create an app in order to use the Twitter API, and Twitter data includes a ``user-agent'' that identifies the app. Some user-agent values correspond to Twitter native apps (\texttt{TweetDeck}, \texttt{Twitter for Advertisers}, \texttt{Twitter for Advertisers (legacy)}, \texttt{Twitter for Android}, \texttt{Twitter for iPad}, \texttt{Twitter for iPhone}, \texttt{Twitter for Mac}, \texttt{Twitter Media Studio}, \texttt{Twitter Web App}, and \texttt{Twitter Web Client}). While software could in principle be written to control native apps, we assume the vast majority of these apps are operated manually. Similarly, we assume non-native apps indicate the use of the Twitter API and thus likely automation, even though some could be operated manually.
The entropy and automation scores are averaged across the accounts in each cluster. 
The clusters in Fig.~\ref{fig:bloc_coord_composite} are well separated along the automation axis, suggesting  a robust distinction between human and bot accounts.

We manually inspected the clusters in Fig.~\ref{fig:bloc_coord_composite} to describe the dominant behaviors, summarized in the groups below. Each cluster number has a suffix indicating the month when it was observed. All the clusters in each group have the same color in Fig.~\ref{fig:bloc_coord_composite}.

\begin{itemize}
    \item \textbf{Giant connected component (blue):} \textbf{Cluster 3-Sep} includes accounts with low automation scores and high variety of behaviors. These are likely legitimate users who mostly retweet and occasionally post tweets, with normal pauses. Similar large components were present on each month. 

    \item \textbf{Vaccine availability/appointment bots (orange):} \textbf{Cluster 12-Apr} includes 12 self-identified bot accounts that track the availability of vaccines and appointments in various US cities, such as \texttt{@DCVaxAlerts} and \texttt{@FindAVac\_Austin}. These accounts posted messages such as ``New available appointments detected! -- Provider: CVS Pharmacy -- City: Alamo Heights -- Registration link: \url{www.cvs.com/immunizations/covid-19-vaccine}...'' They created long bursts of tweets consisting mostly of URLs and text. Overall, these accounts posted the most content. Similarly, \textbf{Cluster 17-Jan} includes two vaccine appointment bots (\texttt{@kcvaccinewatch} and \texttt{@stlvaccinewatch}) that created tweet threads. \textbf{Cluster 13-Jul} includes \texttt{@CovidvaxDEL}, a vaccine appointment status bot for New Delhi, India; and \texttt{@ncovtrack}, a bot that posted vaccine statistics for various countries. 
    
    \item \textbf{News posting accounts (green):} \textbf{Clusters 14-Apr}, \textbf{16-Jan}, \textbf{20-Apr} and \textbf{22-Feb} include many accounts that mostly post tweets linking to news websites hourly, such as \texttt{@canada4news} and \texttt{@HindustanTimes}. Some accounts are owned by international news organizations such as \texttt{@Independent} and \texttt{@guardian}. 
    
    \item \textbf{Content amplifying, likely bot accounts (purple):} \textbf{Cluster 4-May} includes a pair of accounts that create no content; they retweet mostly the same tweets repeatedly. \textbf{Cluster 19-May} includes self-identified bots created by the same self-identified developer. These bots, \texttt{@EdinburghWatch} and \texttt{Glasgow\_Watch}, retweet random content from Glasgow and Edinburgh, respectively.
    
    \item \textbf{Misinformation sharing and local news accounts (white):} \textbf{Cluster 24-Feb} includes \texttt{@USSANews} owned by \texttt{ussanews.com}, a misinformation website according to \texttt{factcheck.org}. This account posted links with headlines such as: ``31 Reasons Why I Won't Take the Vaccine.'' The same cluster includes \texttt{@abc7newsbayarea}, the account of a legitimate local news organization. Both accounts mostly post multiple tweets with images separated by pauses under an hour.
    
    \item \textbf{Spam bots (red):} \textbf{Clusters 10-Mar}, \textbf{11-Apr}, and \textbf{23-Aug} include accounts that post repeated content. The accounts in \textbf{Cluster 10-Mar} repeatedly linked to their respective blogs with exactly seven or thirteen hashtags. \textbf{Cluster 11-Apr} posted messages soliciting others to follow a specified account. The two accounts in \textbf{Cluster 23-Aug} posted the same pro-vaccine messages repeatedly, 133 and 72 times respectively.
    
    \item \textbf{Coordinated bots (black):} The three accounts in \textbf{Cluster 21-May} created no content; they retweeted the same account exactly 1,004 times each. During the first week of May 2021, the first 44 characters of their BLOC strings matched. Similarly, accounts in \textbf{Cluster 15-May} did not create content but always retweeted the same collection of multiple business accounts advertising various merchandise. \textbf{Cluster 18-Mar} includes a pair of accounts that retweeted one another 313 times.
    
    \item \textbf{Various low automation accounts with different stances on vaccine (yellow):} Finally, Fig. \ref{fig:bloc_coord_composite} also features clusters of accounts with pro-vaccine (\textbf{Clusters 1-May} and \textbf{2-Jan}), anti-vaccine (\textbf{Clusters 5-Mar}, \textbf{6-Apr}, \textbf{7-Mar}, and \textbf{8-May}), or a mixture of both sentiments (\textbf{Cluster 9-Jun}).
\end{itemize}

\section{Evaluation}
\label{sec:eval}

In this section we evaluate the performance of BLOC models on bot and coordination detection tasks on Twitter. 
BLOC code and datasets used in our experiments are 
available~\cite{anwala_bloc_code}.

\subsection{Bot detection}

The bot detection task involves separating accounts that are likely operated by human users from accounts that are likely automated. This is a challenging task, as behaviors of both classes of accounts are heterogeneous and time-evolving. 

\subsubsection{Methods}
\label{subsec:individual_bot_baseline}

The BLOC language parameters used for the evaluation are as follows: $p_1 = 1$ minute, $p_2 = f_2(\Delta)$, and $p_4 = \text{bi-gram}$ (Table \ref{tab:bloc_parameters_states}). The other parameters are not applicable to bi-gram tokenization. 
We extracted BLOC \textit{action} and \textit{content} bi-grams for each annotated Twitter account. 
This resulted in a set of 197 bi-grams. 
These bi-grams can be used as features within any machine learning model. 
We obtained TF-IDF feature vectors for each account and used them to train a random-forest classifier.

We compared the performance of BLOC to three baseline models: Botometer-V4 (the current version of Botometer at the time of writing) \cite{sayyadiharikandeh2020detection} and two DNA-based methods, namely DDNA~\cite{cresci2017social,cresci2017exploiting} and DNA-influenced~\cite{gilmary2022dna}. The latter were selected because they share some similarities with BLOC.

Botometer-V4 utilizes 1,161 different features that can be grouped into six categories that focus on different account characteristics. For example, user profile features are extracted from the user profile, like the numbers of friends and followers. Temporal features measure temporal patterns of the posts, such as frequency and times of day. In the deployed system, different classifiers in an ensemble are trained on different accounts types, and then these classifiers vote to obtain the final bot score~\cite{sayyadiharikandeh2020detection}.
Here instead, to compare the representation power of BLOC vs.~Botometer features with all other things being equal, we trained a single random-forest classifier with the same features used to train Botometer-V4.

Digital DNA classifies accounts as bots if they share long sequences of symbols representing actions and content. Cresci et al.~\cite{cresci2017exploiting} provided their Python code~\cite{twitter_dna_code}, which wraps a C implementation for the Longest Common Substring (LCS) algorithm. We modified the code to implement the method described by the authors. The method yields a maximal common substring length from the training data. This length is then used to determine a set of accounts in the test data that share a maximal common substring of the same length. These accounts are classified as bots. We finally apply cross-validation to evaluate the classifier.

The DNA-influenced bot classifier is based on the rationale that bot accounts are more likely to be similar to each other, compared to human accounts. The method relies on a formula to calculate a probability distribution for a given string, and on the \textit{symmetrized KL divergence} to calculate the distance between the probability distributions associated with two strings~\cite{yu2011dna}. In this way, the method calculates the distance between the DDNA strings corresponding to two accounts~\cite{gilmary2022dna}. To implement this method, we partitioned the bot accounts in the training dataset into groups of 50, similar to Gilmary et al.~\cite{gilmary2022dna}. For each group, we calculated the average distances across all pairs of accounts in the group. The maximum average distance across all the groups was then used as a decision threshold: any two accounts in the test dataset were classified as bots if their distance was less than or equal to the decision threshold.

\subsubsection{Datasets}
\label{subsec:eval_dataset}

\begin{table}
\centering
\caption{Annotated datasets used in our bot detection evaluation. For each dataset, we report the reference describing it and the number of accounts that are still active at the time of the present evaluation.}
\begin{tabular}{lcrr}
\hline
\textbf{Dataset} & \textbf{Ref.} & \textbf{\# Bots} & \textbf{\# Humans} \\ \hline
astroturf-20           & \cite{sayyadiharikandeh2020detection} & 505 & 0             \\
botometer-feedback-19  & \cite{yang2019arming} & 123             & 364               \\
botwiki-19             & \cite{yang2020scalable} & 695             & 0               \\
celebrity-19           & \cite{yang2019arming}  & 0            & 20,911                   \\
cresci-17              & \cite{cresci2017paradigm}  & 5,812    & 2,744               \\
cresci-rtbust-19       & \cite{mazza2019rtbust} & 352             & 340               \\
cresci-stock-18        & \cite{cresci2018fake}  & 6,926           & 6,155         \\
gilani-17              & \cite{gilani2017bots} & 1,058           & 1,381             \\
midterm-18             & \cite{yang2020scalable} & 0               & 7,409             \\
political-bots-19      & \cite{yang2019arming} & 62              & 0                 \\
pronbots-19            & \cite{yang2019arming} & 14,867           & 0                 \\
varol-17               & \cite{varol2017online} & 728             & 1,483             \\
vendor-purchased-19    & \cite{yang2019arming} & 928           & 0                 \\
verified-19            & \cite{yang2020scalable} & 0               & 1,986              \\ \hline
\textbf{Total}                  & - & 32,056           & 42,773              \\ \hline
\end{tabular}
\label{tab:bot_eval_dataset}
\end{table}

Our evaluation datasets (Table~\ref{tab:bot_eval_dataset}) consist of 32,056 Twitter accounts labeled as bots and 42,773 accounts labeled as humans, all selected from the bot repository.\footnote{\url{botometer.osome.iu.edu/bot-repository}}
These accounts were collected and labeled by multiple researchers between 2017--2019~\cite{yang2019arming}. 
To eliminate a potential bias in the comparative analysis that might result from the class imbalance, we took the union of all datasets but used a random sample of 32,056 accounts from the majority class (humans). 

\subsubsection{Results}

We evaluated BLOC, Botometer, three variants of Digital DNA (\textit{b3\_type}, \textit{b3\_content}, and \textit{b6\_content})~\cite{cresci2017exploiting}, and DNA-influenced by predicting bot and human labels, all on the same annotated dataset in Table~\ref{tab:bot_eval_dataset}. We computed precision, recall, and $F_1$ from 5-fold cross validation.

\begin{table}
\centering
\caption{Precision, recall, and $F_1$ for different bot classifiers using 5-fold cross-validation, along with numbers of features. The best values for each metric are shown in bold. DNA-influenced classifiers produced recall of 1.0 because they always predicted that all account were bots.}
\begin{tabular}{lrrrr}
\hline
\textbf{Model}                      & \textbf{Precision} & \textbf{Recall} & \textbf{$F_1$} & \textbf{Features} \\ \hline
    
    BLOC                                & .899          & .884        &  .892 & \textbf{197}  \\ 
    Botometer                           & \textbf{.929}          & .914        &  \textbf{.921} & 1,160\\ 
    DNA-influenced                      & .499          & \textbf{1.000}    &  .666 & --            \\ 
    Digital DNA ($b3\_type$)            & .796          & .529        &  .636 & --            \\ 
    Digital DNA ($b3\_content$)         & .866          & .183        &  .303 & --            \\ 
    Digital DNA  ($b6\_content$)        & .868          & .187        &  .308 & --            \\ \hline
\end{tabular}
\label{tab:bloc_vs_rest_bot_detection}
\end{table}

As reported in Table~\ref{tab:bloc_vs_rest_bot_detection}, Botometer-V4 slightly outperformed BLOC on the $F_1$ metric. However, BLOC used significantly fewer features. DNA-influenced outperformed Digital DNA, even though it labeled \textit{all} accounts as bots.

\subsection{Coordination detection}
\label{subsec:id_coord_detect_info_ops}

Multiple nation states utilize social media for information operations that target their citizens, foreign nationals, organizations, etc. Twitter defines information operations as a form of platform abuse, which involves artificial amplification or suppression of information or behavior that manipulates or disrupts the user experience.\footnote{\url{help.twitter.com/en/rules-and-policies/platform-manipulation}} Twitter deletes the public tweets of accounts engaged in information operations, but publishes datasets containing these tweets. 

Let us use the term \textit{driver} to refer to an account engaged in some information operation. Drivers may employ tactics such as spamming, impersonation, obfuscation, and/or targeting of individuals or communities. We consider all these behaviors \textit{coordinated} but do not distinguish among them. Our task is to separate the drivers from regular (control) accounts tweeting about the same topics. 

\subsubsection{Methods}

Coordination detection is based on unsupervised learning, namely, identifying accounts with suspiciously similar behaviors. BLOC words express behavioral traits. We generated TF-IDF vectors as described in Section \ref{subsec:individual_bot_baseline} and then calculated the similarity between two accounts via the cosine between their two vectors.

We compared BLOC to three baseline methods, which make different assumptions about the behavioral traits that may be shared among coordinated accounts~\cite{pacheco2020uncovering}: \textit{hashtag sequences} (Hash), \textit{activity} (Activity), and \textit{co-retweet} (CoRT).  
The hashtag baseline method identifies coordinating accounts by finding those that mostly use the same sequences of hashtags (e.g., the same hashtag 5-grams). The activity method looks for accounts that are synchronized in the times when they post tweets: accounts that often tweet or retweet within the same time window are considered suspicious. Similar to Pacheco et. al.~\cite{pacheco2020uncovering}, we considered accounts that consistently posted tweets within 30-minutes from one another to be suspicious. The co-retweet method identifies coordinating accounts by finding those that mostly retweet the same sets of tweets. 
We generated TF-IDF vectors of hashtag 5-grams, of activity time intervals, and of retweeted tweet IDs, as described by Pacheco et. al.~\cite{pacheco2020uncovering}. For all baselines, the cosine between TF-IDF vectors was used to calculate similarity.

We also evaluated a \textit{combined} method. For a pair of accounts, the combined method takes the maximum among four cosine similarity values computed with BLOC and the three baselines. 

We built $k$-nearest-neighbor (KNN) classifiers using $k = 1, \dots, 10$ to compare the five methods. We report the maximum $F_1$ obtained across the $k$ values. 

\subsubsection{Datasets}

\begin{table}
\centering
\caption{Selected information operation. We list life spans, the number of weeks used for evaluation (since the start of the information operations), and the counts of drivers and control accounts active during the evaluation weeks. Note that the evaluation weeks are not necessarily contiguous.}
\begin{tabular}{lcrrr}
\hline
\textbf{Information Op.} & \textbf{Life span} & \textbf{Eval. week} & \textbf{\# Drivers} & \textbf{\# Control} \\ \hline
Armenia & 2014 -- 2020 & 4 & 27 & 1,462 \\
Bangladesh & 2010 -- 2018 & 8 & 10 & 929 \\
Catalonia & 2011 -- 2019 & 20 & 14 & 906 \\
China\_1 & 2008 -- 2019 & 20 & 11 & 905 \\
China\_2 & 2008 -- 2019 & 44 & 53 & 4,019 \\
China\_3 & 2021 & 4 & 95 & 165 \\
China\_4 & 2020 -- 2021 & 18 & 1,623 & 4,681 \\
China\_5 & 2021 & 10 & 1,247 & 4,890 \\
Cuba & 2010 -- 2020 & 26 & 11 & 3,415 \\
Ecuador & 2010 -- 2019 & 22 & 10 & 1,767 \\
Egypt\_UAE & 2012 -- 2019 & 40 & 59 & 350 \\
Ghana\_Nigeria & 2014 -- 2020 & 28 & 53 & 1,102 \\
Iran\_1 & 2012 -- 2018 & 32 & 23 & 824 \\
Iran\_2 & 2010 -- 2015 & 22 & 11 & 407 \\
Iran\_3 & 2011 -- 2020 & 20 & 14 & 791 \\
Iran\_4 & 2014 -- 2019 & 38 & 13 & 2,088 \\
Iran\_5 & 2013 -- 2019 & 28 & 14 & 701 \\
Iran\_6 & 2020 -- 2020 & 8 & 104 & 1,247 \\
Iran\_7 & 2010 -- 2020 & 52 & 16 & 11,842 \\
Mexico\_1 & 2019 -- 2019 & 18 & 119 & 2,097 \\
Mexico\_2 & 2020 -- 2021 & 44 & 240 & 6,340 \\
Qatar & 2013 -- 2020 & 18 & 11 & 7,393 \\
Russia\_1 & 2009 -- 2018 & 4 & 12 & 498 \\
Russia\_2 & 2011 -- 2018 & 2 & 10 & 168 \\
Russia\_3 & 2009 -- 2020 & 6 & 10 & 5,200 \\
Russia\_4 & 2014 -- 2020 & 24 & 21 & 2,973 \\
Spain & 2019 -- 2019 & 8 & 215 & 1,681 \\
Thailand & 2018 -- 2020 & 8 & 166 & 1,133 \\
UAE & 2011 -- 2019 & 34 & 17 & 1,662 \\
Uganda\_1 & 2019 -- 2020 & 54 & 124 & 11,323 \\
Uganda\_2 & 2020 -- 2021 & 54 & 342 & 10,526 \\
Venezuela\_1 & 2010 -- 2018 & 28 & 48 & 891 \\
Venezuela\_2 & 2015 -- 2018 & 8 & 71 & 995 \\
Venezuela\_3 & 2012 -- 2019 & 16 & 17 & 1,412 \\
Venezuela\_4 & 2020 -- 2021 & 54 & 139 & 12,365 \\
Venezuela\_5 & 2021 & 22 & 249 & 5,447 \\
 \hline
\textbf{Total}                  & &  & 5,219           & 114,595              \\ \hline
\end{tabular}
\label{tab:info_ops_dataset_first_active_years}
\end{table}

Twitter published over 141 information operation datasets~\cite{twitter_info_ops}. These datasets include tweets by drivers across 21 countries, during different time periods between 2008 and 2021. 
To ensure a fair assessment of the classifiers for detecting information operation drivers, we built control datasets that include tweets by accounts not engaged in information operations, but who posted about the same topics around the same time. For each information operation, we extracted all the hashtags used by the drivers. Then we used these hashtags as queries to Twitter's academic search API,\footnote{\url{developer.twitter.com/en/products/twitter-api/academic-research}} which does not impose date restrictions. We extracted accounts that posted tweets on the same dates and with the same hashtags as the drivers. Finally, for each of these accounts, we reconstructed their timelines by extracting a maximum of 100 tweets posted on the same dates as the drivers. 
We were able to create control datasets for 36 information operations, as shown in Table~\ref{tab:info_ops_dataset_first_active_years}. 
These represent 18 of the countries and the entire time period. 

Some information operations lasted a few months (e.g., China\_3 in Table~\ref{tab:info_ops_dataset_first_active_years}), others over five years (e.g., Iran\_7 in Table~\ref{tab:info_ops_dataset_first_active_years}). Therefore, we could run the experiment of detecting drivers for different time periods (e.g., first year, last year, all years). From the perspective of mitigation, we followed the principle that it is desirable to detect drivers as early as possible, with as little information (tweets) as possible. We believe it is more difficult to detect drivers early, since sufficient tweets with coordination signals might be absent. 

Based on the above principle, we ran each experiment by incrementally adding two weeks of data until the end of the first year in which at least 10 drivers were observed, or the end of the campaign --- whichever occurred first. In other words, the first instance of our experiment was run on two weeks of data, the second on four weeks of data, and so on. 
The use of increasing evaluation intervals is meant to explore how accuracy depends on the amount of data accumulated. 
For each coordination detection method, we generated vectors corresponding to all driver and control accounts active in each information operation and evaluation interval. 
Table~\ref{tab:info_ops_dataset_first_active_years} reports on the full evaluation periods and numbers of driver and control accounts in our datasets.

\subsubsection{Results}

\begin{figure}
  \centering
  \includegraphics[width=\linewidth]{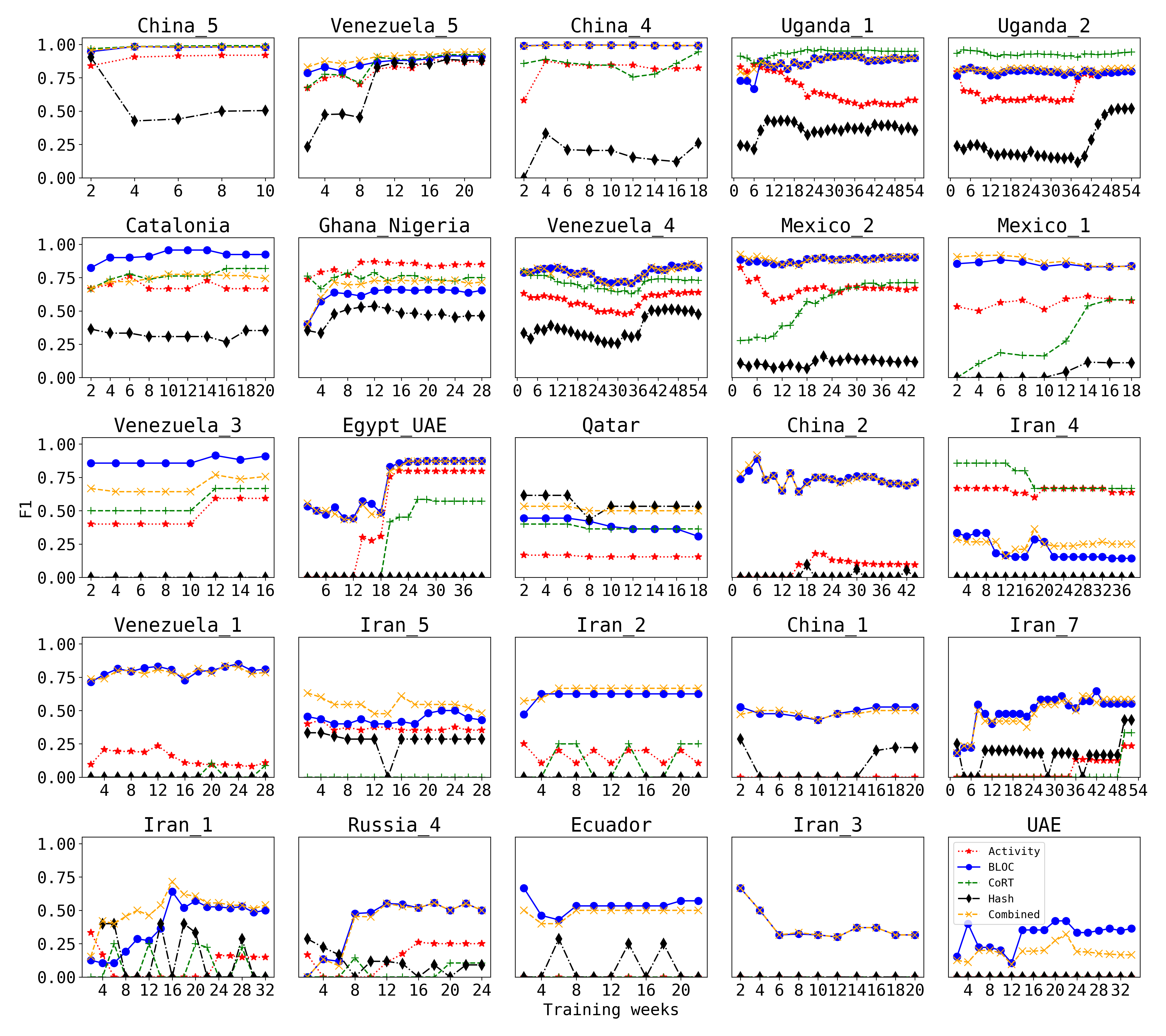}
  \caption{$F_1$ scores of the best-performing classifiers for detecting information operation drivers for the subset of campaigns with at least 10 weeks worth of data. The weeks displayed on the x-axis represent those in which the drivers were active (evaluation weeks); they are not necessarily contiguous. The plots are ordered in descending order of $F_1$ score calculated at week 10 using the combined method.}
  \label{fig:k_first_active_years_info_ops_bloc_v_rest_knn}
\end{figure}

Fig.~\ref{fig:k_first_active_years_info_ops_bloc_v_rest_knn} plots the $F_1$ values of the best-performing classifiers for a subset of information operations. The best KNN classifier is the one with the $k$ value ($k = 1, \dots, 10$) yielding the maximum $F_1$. The x-axis for each plot represents the number of evaluation weeks, while the y-axis represents the $F_1$ score of the best classifier. The information operations are ordered in descending order of their respective \textit{combined F1@Week 10} score, to capture the difficulty of detecting their drivers. The \textit{combined F1@Week 10} score of an information operation is the $F_1$ score calculated with 10 weeks worth of data (\textit{F1@Week 10}) using the combined method. Table~\ref{tab:info_ops_eval_result_first_years_active} outlines the \textit{F1@Week 10} scores for all information operations.

\begin{table}
\centering
\caption{$F_1$ scores of BLOC and baseline classifiers for the detection of information operation drivers, calculated with data from the first 10 weeks of each campaign (\textit{F1@Week 10}). For campaigns with less than 10 weeks of data, the entire dataset was used. Information operations are sorted by the $F_1$ score of the combined method (\textit{combined F1@Week 10}). The best method for each campaign is shown in bold. Note that $F_1=0$ when the similarity signal used by a classifier cannot be observed in the behavior of a particular campaign's drivers.
No co-retweets were observed between any pairs of drivers in China\_1.
}
\begin{tabular}{lccccc}
\hline
\textbf{Information Op.} & \textbf{BLOC} & \textbf{Activity} & \textbf{CoRT} & \textbf{Hash} & \textbf{Combined} \\ \hline
China\_3 & \textbf{.995} & .968 & .182 & .973 & .995 \\
China\_4 & \textbf{.996} & .846 & .844 & .205 & .994 \\
China\_5 & .981 & .919 & \textbf{.991} & .504 & .980 \\
Iran\_6 & .961 & \textbf{.986} & .978 & .000 & .976 \\
Venezuela\_2 & \textbf{.936} & .864 & .000 & .000 & .929 \\
Venezuela\_5 & .869 & .815 & \textbf{.903} & .830 & .910 \\
Spain & .876 & \textbf{.936} & .892 & .859 & .904 \\
Mexico\_2 & \textbf{.851} & .569 & .312 & .071 & .874 \\
Mexico\_1 & \textbf{.833} & .511 & .163 & .000 & .859 \\
Uganda\_1 & .850 & .807 & \textbf{.898} & .432 & .850 \\
Thailand & .769 & \textbf{.815} & .808 & .226 & .832 \\
Uganda\_2 & .800 & .574 & \textbf{.940} & .227 & .807 \\
Venezuela\_4 & \textbf{.819} & .603 & .754 & .391 & .779 \\
Venezuela\_1 & \textbf{.820} & .188 & .000 & .000 & .776 \\
Catalonia & \textbf{.957} & .667 & .762 & .308 & .774 \\
China\_2 & \textbf{.762} & .000 & .000 & .000 & .762 \\
Armenia & .760 & .875 & \textbf{.895} & .851 & .760 \\
Ghana\_Nigeria & .612 & \textbf{.866} & .741 & .528 & .700 \\
Russia\_1 & \textbf{.667} & .000 & .000 & .000 & .667 \\
Iran\_2 & \textbf{.625} & .200 & .000 & .000 & .667 \\
Venezuela\_3 & \textbf{.857} & .400 & .500 & .000 & .643 \\
Iran\_5 & \textbf{.435} & .353 & .000 & .286 & .545 \\
Qatar & .381 & .154 & .364 & \textbf{.533} & .500 \\
Russia\_3 & \textbf{.500} & .000 & .000 & .000 & .500 \\
Ecuador & \textbf{.533} & .000 & .000 & .000 & .500 \\
Iran\_1 & \textbf{.286} & .000 & .000 & .000 & .500 \\
Russia\_4 & \textbf{.483} & .000 & .000 & .118 & .452 \\
Egypt\_UAE & \textbf{.444} & .000 & .000 & .000 & .435 \\
China\_1 & \textbf{.429} & .000 &  & .000 & .429 \\
Iran\_7 & \textbf{.476} & .000 & .000 & .200 & .421 \\
Bangladesh & \textbf{.600} & .316 & .333 & .286 & .345 \\
Iran\_3 & \textbf{.316} & .000 & .000 & .000 & .316 \\
Iran\_4 & .182 & .667 & \textbf{.857} & .000 & .267 \\
Russia\_2 & \textbf{.200} & .000 & \textbf{.200} & .000 & .211 \\
UAE & \textbf{.200} & .000 & .000 & .000 & .182 \\
 \hline
\end{tabular}
\label{tab:info_ops_eval_result_first_years_active}
\end{table}

\begin{figure}
    \centering
    \includegraphics[width=0.7\textwidth]{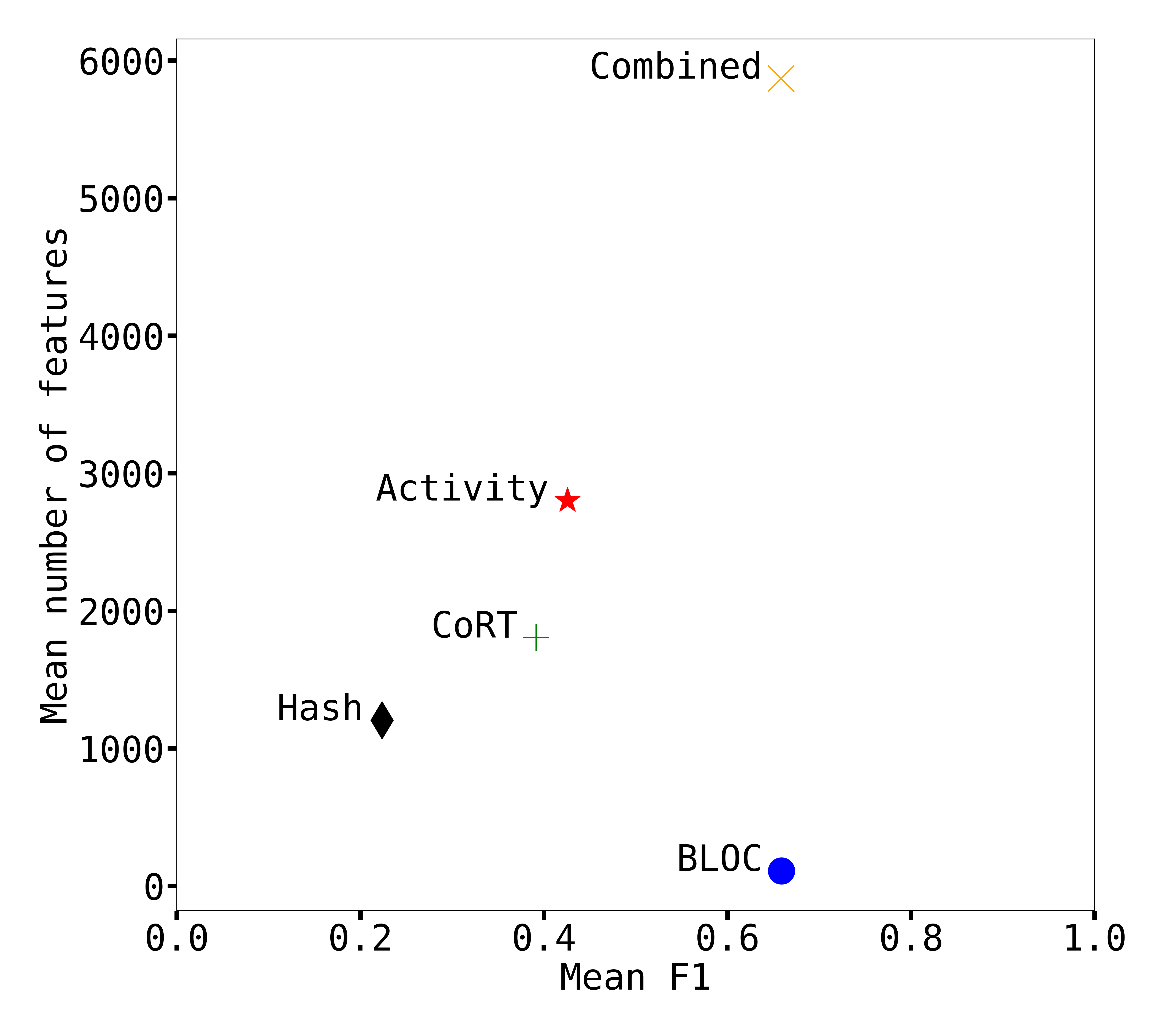}
   
    \caption{Mean number of features vs.~mean $F_1$ of BLOC and four baseline classifiers for detecting drivers across information operations, estimated with data from the first 10 weeks of each information operation's lifespan.}
    \label{fig:bloc_knn_models_eval_first_active}
\end{figure}

According to Fig.~\ref{fig:k_first_active_years_info_ops_bloc_v_rest_knn} and Table~\ref{tab:info_ops_eval_result_first_years_active}, BLOC outperforms the baselines in most campaigns. The drivers from information operations originating from China (e.g., China\_4 and China\_5) were the easiest to detect; the $F_1$ scores for all coordination detection methods except Hash were above 0.9. The hardest drivers to detect were those from the UAE information operation. 
We also note in Fig.~\ref{fig:k_first_active_years_info_ops_bloc_v_rest_knn} that in some campaigns (Venezuela\_4, Venezuale\_3, and Egypt\_UAE), the accuracy of different methods improves in a correlated fashion with more training data. This suggests that drivers display multiple coordination signals simultaneously.  
Yet, having more data does not necessarily imply higher accuracy in detecting drivers. In several campaigns there is no clear temporal trend, and in a few cases (e.g., Iran\_4 and Iran\_3) adding more data hinders detection. This suggest that drivers may change their behaviors and become harder to detect as a result. 

Fig.~\ref{fig:bloc_knn_models_eval_first_active} compares the performance of BLOC and the three baseline coordination detection methods. 
The x-axis represents the mean $F_1$ and the y-axis represents the mean number of features of all classifiers, across all information operations. Both values were calculated with data from the first 10 weeks of the information operations. The BLOC classifiers outperformed all baselines in the coordination detection task with a mean $F_1=0.659$ with the least number of features (108). The combined classifiers had a similar mean $F_1=0.658$, but employed the largest number of features (5,869).

\section{Discussion}

In response to the far-reaching threats posed by influence operations on social media, researchers developed methods that target specific kinds of malicious behaviors. The effectiveness of some of these --- which mostly depend on hand-crafted features --- is however temporary since malicious actors evolve their tactics to evade detection. In this paper, we proposed BLOC, a general language that represents the behavior of social media users irrespective of class (e.g., bot or human) or intent (e.g., benign or malicious). BLOC words map to features derived in an unsupervised manner. We note that BLOC does not make feature engineering irrelevant, in fact one could engineer features using BLOC. 

Although BLOC is platform-agnostic, we demonstrated its flexibility through two real-world applications on Twitter. 
In the bot detection task, BLOC performed better than similar methods (Digital DNA and DNA-influenced) and comparably to a state-of-the-art method (Botometer-V4), with a much lower number of features. 
This indicates that the BLOC action and content alphabets provide useful signals in discriminating between automated and human accounts across a variety of datasets.

In the coordination detection task to identify the drivers of information operations during the early stages of their life span, BLOC outperformed baseline methods. 
The performance of all classifiers varied across information operations, which highlights the heterogeneity of the driver behaviors. This is consistent with Twitter's reports, which reveal that drivers include humans, automated accounts, coordinating accounts, and so on~\cite{twitter_info_ops_2020, twitter_info_ops_2021}. So it comes as no surprise that the drivers of some information operations are easier than others to detect. 

We also compared the performance of all classifiers on information operation datasets extracted from the \textit{last} weeks of the life spans of the drivers --- right before they were detected by Twitter.
The average $F_1$ scores of all classifiers increased significantly (by 25--101\%), 
suggesting that Twitter detected the drivers when their behaviors became more conspicuous. 
The activity method slightly outperformed BLOC when the evaluation was run during the last weeks of a campaign, with mean $F_1=0.855$ vs.~0.824, albeit using 1,680 vs.~116 features. This suggests that synchronization is a strong signal for platforms to identify coordinated campaigns.
For example, the activity method failed to identify the drivers of the Egypt\_UAE information operation based on early data ($F_1=0$), but succeeded at the end ($F_1=1$). By contrast, BLOC achieved $F_1=0.444$ based on early data and $F_1=0.978$ at the end. 

Collectively, these results indicate that BLOC models are versatile, effective, efficient, and applicable to multiple tasks. The development of a new technique such as BLOC for identifying malicious behavior could however trigger changes in tactics by ``puppet masters'' to evade detection. 
We argue that even though this is possible, the introduction of BLOC could raise the bar for malicious social media accounts to appear authentic, discounting the benefits of automated tactics such as flooding and coordination.

\section{Acknowledgments}

We are grateful to Manita Pote for help with the control dataset and Kristina Lerman for suggesting the use of pause symbols in the BLOC action alphabet. 
This work was supported in part by DARPA (grants W911NF-17-C-0094 and HR001121C0169), Knight Foundation, and Craig Newmark Philanthropies. 
The funders had no role in study design, data collection and analysis, decision to publish, or preparation of the manuscript.

\bibliographystyle{plain}
\bibliography{bloc}

\begin{thebibliography}{10}

\bibitem{allington2020health}
Daniel Allington, Bobby Duffy, Simon Wessely, Nayana Dhavan, and James Rubin.
\newblock Health-protective behaviour, social media usage and conspiracy belief
  during the covid-19 public health emergency.
\newblock {\em Psychological medicine}, 51(10):1763--1769, 2021.

\bibitem{assenmacher2020two}
Dennis Assenmacher, Lena Clever, Janina~Susanne Pohl, Heike Trautmann, and
  Christian Grimme.
\newblock A two-phase framework for detecting manipulation campaigns in social
  media.
\newblock In {\em Intl. Conf. on Human-Computer Interaction (HCI)}, pages
  201--214. Springer, 2020.

\bibitem{benevenuto2009characterizing}
Fabr{\'\i}cio Benevenuto, Tiago Rodrigues, Meeyoung Cha, and Virg{\'\i}lio
  Almeida.
\newblock Characterizing user behavior in online social networks.
\newblock In {\em Proc. ACM SIGCOMM Conf. on Internet Measurement (IMC)}, pages
  49--62, 2009.

\bibitem{beskow2018bot}
David~M Beskow and Kathleen~M Carley.
\newblock {Bot conversations are different: leveraging network metrics for bot
  detection in Twitter}.
\newblock In {\em IEEE/ACM Intl. Conf. on Advances in Social Networks Analysis
  and Mining (ASONAM)}, pages 825--832. IEEE, 2018.

\bibitem{Blondel}
Vincent~D Blondel, Jean-Loup Guillaume, Renaud Lambiotte, and Etienne Lefebvre.
\newblock Fast unfolding of communities in large networks.
\newblock {\em Journal of Statistical Mechanics: Theory and Experiment},
  2008(10):P10008, 2008.

\bibitem{chavoshi2016debot}
Nikan Chavoshi, Hossein Hamooni, and Abdullah Mueen.
\newblock {Debot: Twitter bot detection via warped correlation}.
\newblock In {\em IEEE Intl. Conf. on Data Mining (ICDM)}, pages 817--822,
  2016.

\bibitem{chu2010tweeting}
Zi~Chu, Steven Gianvecchio, Haining Wang, and Sushil Jajodia.
\newblock {Who is tweeting on Twitter: human, bot, or cyborg?}
\newblock In {\em Proc. of Annual Computer Security Applications Conference
  (ACSAC)}, pages 21--30, 2010.

\bibitem{chu2012detecting}
Zi~Chu, Steven Gianvecchio, Haining Wang, and Sushil Jajodia.
\newblock {Detecting automation of Twitter accounts: Are you a human, bot, or
  cyborg?}
\newblock {\em IEEE Transactions on Dependable and Secure Computing},
  9(6):811--824, 2012.

\bibitem{cresci2020decade}
Stefano Cresci.
\newblock A decade of social bot detection.
\newblock {\em Communications of the ACM}, 63(10):72--83, 2020.

\bibitem{cresci2017exploiting}
Stefano Cresci, Roberto Di~Pietro, Marinella Petrocchi, Angelo Spognardi, and
  Maurizio Tesconi.
\newblock {Exploiting digital DNA for the analysis of similarities in Twitter
  behaviours}.
\newblock In {\em IEEE Intl. Conf. on Data Science and Advanced Analytics
  (DSAA)}, pages 686--695. IEEE, 2017.

\bibitem{cresci2017paradigm}
Stefano Cresci, Roberto Di~Pietro, Marinella Petrocchi, Angelo Spognardi, and
  Maurizio Tesconi.
\newblock The paradigm-shift of social spambots: Evidence, theories, and tools
  for the arms race.
\newblock In {\em Proc. of Intl. Conf. Companion on World Wide Web}, pages
  963--972, 2017.

\bibitem{cresci2017social}
Stefano Cresci, Roberto Di~Pietro, Marinella Petrocchi, Angelo Spognardi, and
  Maurizio Tesconi.
\newblock Social fingerprinting: detection of spambot groups through
  dna-inspired behavioral modeling.
\newblock {\em IEEE Transactions on Dependable and Secure Computing},
  15(4):561--576, 2017.

\bibitem{cresci2018fake}
Stefano Cresci, Fabrizio Lillo, Daniele Regoli, Serena Tardelli, and Maurizio
  Tesconi.
\newblock {\$ FAKE: Evidence of spam and bot activity in stock microblogs on
  Twitter}.
\newblock In {\em Proc. Intl. AAAI Conf. on Web and Social Media (ICWSM)},
  2018.

\bibitem{cresci2019cashtag}
Stefano Cresci, Fabrizio Lillo, Daniele Regoli, Serena Tardelli, and Maurizio
  Tesconi.
\newblock {Cashtag piggybacking: Uncovering spam and bot activity in stock
  microblogs on Twitter}.
\newblock {\em ACM Transactions on the Web (TWEB)}, 13(2):1--27, 2019.

\bibitem{davis2016botornot}
Clayton~Allen Davis, Onur Varol, Emilio Ferrara, Alessandro Flammini, and
  Filippo Menczer.
\newblock Botornot: A system to evaluate social bots.
\newblock In {\em Proc. of Intl. Conf. Companion on World Wide Web}, pages
  273--274, 2016.

\bibitem{DeVerna_Pierri_2021}
Matthew~R. DeVerna, Francesco Pierri, Bao~Tran Truong, John Bollenbacher, David
  Axelrod, Niklas Loynes, Christopher Torres-Lugo, Kai-Cheng Yang, Filippo
  Menczer, and John Bryden.
\newblock {CoVaxxy: A Collection of English-Language Twitter Posts About
  COVID-19 Vaccines}.
\newblock {\em Proc. Intl. AAAI Conf. on Web and Social Media (ICWSM)},
  15(1):992--999, 2021.

\bibitem{fazil2020socialbots}
Mohd Fazil and Muhammad Abulaish.
\newblock {A socialbots analysis-driven graph-based approach for identifying
  coordinated campaigns in Twitter}.
\newblock {\em Journal of Intelligent \& Fuzzy Systems}, 38(3):2961--2977,
  2020.

\bibitem{ferrara2016rise}
Emilio Ferrara, Onur Varol, Clayton Davis, Filippo Menczer, and Alessandro
  Flammini.
\newblock The rise of social bots.
\newblock {\em Communications of the ACM}, 59(7):96--104, 2016.

\bibitem{ghosh2011entropy}
Rumi Ghosh, Tawan Surachawala, and Kristina Lerman.
\newblock {Entropy-based Classification of Retweeting Activity on Twitter}.
\newblock In {\em Proc. of KDD Workshop on Social Network Analysis (SNA-KDD)},
  2011.

\bibitem{giglietto2020coordinated}
Fabio Giglietto, Nicola Righetti, Luca Rossi, and Giada Marino.
\newblock Coordinated link sharing behavior as a signal to surface sources of
  problematic information on facebook.
\newblock In {\em Intl. Conf. on Social Media and Society}, pages 85--91, 2020.

\bibitem{giglietto2020takes}
Fabio Giglietto, Nicola Righetti, Luca Rossi, and Giada Marino.
\newblock It takes a village to manipulate the media: coordinated link sharing
  behavior during 2018 and 2019 italian elections.
\newblock {\em Information, Communication \& Society}, 23(6):867--891, 2020.

\bibitem{gilani2017bots}
Zafar Gilani, Reza Farahbakhsh, Gareth Tyson, Liang Wang, and Jon Crowcroft.
\newblock {Of bots and humans (on Twitter}).
\newblock In {\em Proc. of Intl. Conf. on Advances in Social Networks Analysis
  and Mining (ASONAM)}, pages 349--354. ACM, 2017.

\bibitem{gilmary2022dna}
Rosario Gilmary, Akila Venkatesan, Govindasamy Vaiyapuri, and Deepikashini
  Balamurali.
\newblock {DNA-influenced automated behavior detection on Twitter through
  relative entropy}.
\newblock {\em Scientific Reports}, 12(1):1--12, 2022.

\bibitem{grinberg2019fake}
Nir Grinberg, Kenneth Joseph, Lisa Friedland, Briony Swire-Thompson, and David
  Lazer.
\newblock {Fake news on Twitter during the 2016 US presidential election}.
\newblock {\em Science}, 363(6425):374--378, 2019.

\bibitem{he2014identifying}
Su~He, Hui Wang, and Zhi~Hong Jiang.
\newblock {Identifying user behavior on Twitter based on multi-scale entropy}.
\newblock In {\em Proc. of IEEE Intl. Conf. on Security, Pattern Analysis, and
  Cybernetics (SPAC)}, pages 381--384. IEEE, 2014.

\bibitem{jurafskyspeech}
Daniel Jurafsky and James~H Martin.
\newblock {\em Speech and Language Processing: An Introduction to Natural
  Language Processing, Computational Linguistics, and Speech Recognition}.
\newblock Prentice Hall, 2nd edition, 2018.

\bibitem{keller2017manipulate}
Franziska Keller, David Schoch, Sebastian Stier, and JungHwan Yang.
\newblock How to manipulate social media: Analyzing political astroturfing
  using ground truth data from south korea.
\newblock In {\em Proc. of Intl. AAAI Conf. on Web and Social Media (ICWSM)},
  2017.

\bibitem{keller2020political}
Franziska~B Keller, David Schoch, Sebastian Stier, and JungHwan Yang.
\newblock {Political astroturfing on Twitter: How to coordinate a
  disinformation campaign}.
\newblock {\em Political Communication}, 37(2):256--280, 2020.

\bibitem{Lazer-fake-news-2018}
David Lazer, Matthew Baum, Yochai Benkler, Adam Berinsky, Kelly Greenhill,
  Filippo Menczer, Miriam Metzger, Brendan Nyhan, Gordon Pennycook, David
  Rothschild, Michael Schudson, Steven Sloman, Cass Sunstein, Emily Thorson,
  Duncan Watts, and Jonathan Zittrain.
\newblock The science of fake news.
\newblock {\em Science}, 359(6380):1094--1096, 2018.

\bibitem{lee2011seven}
Kyumin Lee, Brian~David Eoff, and James Caverlee.
\newblock {Seven months with the devils: A long-term study of content polluters
  on Twitter}.
\newblock In {\em Proc. Intl. AAAI Conf. on Web and Social Media (ICWSM)},
  2011.

\bibitem{magelinski2021synchronized}
Thomas Magelinski, Lynnette Hui~Xian Ng, and Kathleen~M Carley.
\newblock A synchronized action framework for responsible detection of
  coordination on social media.
\newblock Preprint 2105.07454, arXiv, 2021.

\bibitem{maia2008identifying}
Marcelo Maia, Jussara Almeida, and Virg{\'\i}lio Almeida.
\newblock Identifying user behavior in online social networks.
\newblock In {\em Proc. Workshop on Social Network Systems}, pages 1--6, 2008.

\bibitem{fisher2013syrian}
{Max Fisher}.
\newblock {Syrian hackers claim AP hack that tipped stock market by \$136
  billion. Is it terrorism}.
\newblock \url{https://archive.ph/VJzwk}, 2013.
\newblock Accessed: 2022-04-12.

\bibitem{mazza2019rtbust}
Michele Mazza, Stefano Cresci, Marco Avvenuti, Walter Quattrociocchi, and
  Maurizio Tesconi.
\newblock {Rtbust: Exploiting temporal patterns for botnet detection on
  Twitter}.
\newblock In {\em Proc. of ACM Conference on Web Science (WebSci)}, pages
  183--192, 2019.

\bibitem{mikolov2013distributed}
Tomas Mikolov, Ilya Sutskever, Kai Chen, Greg~S Corrado, and Jeff Dean.
\newblock Distributed representations of words and phrases and their
  compositionality.
\newblock In C.J. Burges, L.~Bottou, M.~Welling, Z.~Ghahramani, and K.Q.
  Weinberger, editors, {\em Advances in Neural Information Processing Systems},
  volume~26, 2013.

\bibitem{nightingale2022ai}
Sophie~J. Nightingale and Hany Farid.
\newblock {AI}-synthesized faces are indistinguishable from real faces and more
  trustworthy.
\newblock {\em Proc. of the National Academy of Sciences}, 119(8):e2120481119,
  2022.

\bibitem{nizzoli2021coordinated}
Leonardo Nizzoli, Serena Tardelli, Marco Avvenuti, Stefano Cresci, and Maurizio
  Tesconi.
\newblock Coordinated behavior on social media in 2019 uk general election.
\newblock In {\em Proc. Intl. AAAI Conf. on Web and Social Media (ICWSM)},
  pages 443--454, 2021.

\bibitem{anwala_bloc_code}
Alexander Nwala, Alessandro Flammini, and Filippo Menczer.
\newblock {A General Language for Modeling Social Media Account Behavior}.
\newblock \url{https://github.com/anwala/general-language-behavior}, 2022.
\newblock Accessed: 2022-10-10.

\bibitem{pacheco2020uncovering}
Diogo Pacheco, Pik-Mai Hui, Christopher Torres-Lugo, Bao~Tran Truong,
  Alessandro Flammini, and Filippo Menczer.
\newblock Uncovering coordinated networks on social media: Methods and case
  studies.
\newblock In {\em Proc. Intl. AAAI Conf. on Web and Social Media (ICWSM)},
  volume~15, pages 455--466, 2021.

\bibitem{covaxxy-misinfo}
Francesco Pierri, Brea Perry, Matthew~R. DeVerna, Kai-Cheng Yang, Alessandro
  Flammini, Filippo Menczer, and John Bryden.
\newblock {Online misinformation is linked to early COVID-19 vaccination
  hesitancy and refusal}.
\newblock {\em Scientific Reports}, 12:5966, 2022.

\bibitem{ratkiewicz2011detecting}
Jacob Ratkiewicz, Michael Conover, Mark Meiss, Bruno Gon{\c{c}}alves,
  Alessandro Flammini, and Filippo Menczer.
\newblock Detecting and tracking political abuse in social media.
\newblock In {\em Proc. Intl. AAAI Conf. on Weblogs and Social Media (ICWSM)},
  2011.

\bibitem{twitter_dna_code}
Bellomo Salvatore, Cresci Stefano, Gagliano Giuseppe, Martella Antonio,
  Spognardi Angelo, and Tesconi Maurizio.
\newblock {Digital DNA Toolbox}.
\newblock \url{https://github.com/WAFI-CNR/ddna-toolbox}, 2019.
\newblock Accessed: 2022-08-15.

\bibitem{sayyadiharikandeh2020detection}
Mohsen Sayyadiharikandeh, Onur Varol, Kai-Cheng Yang, Alessandro Flammini, and
  Filippo Menczer.
\newblock Detection of novel social bots by ensembles of specialized
  classifiers.
\newblock In {\em Proc. of ACM Intl. Conf. on Information \& Knowledge
  Management (CIKM)}, pages 2725--2732, 2020.

\bibitem{schiffrin2017disinformation}
Anya Schiffrin.
\newblock Disinformation and democracy: The internet transformed protest but
  did not improve democracy.
\newblock {\em Journal of International Affairs}, 71(1):117--126, 2017.

\bibitem{sharma2021identifying}
Karishma Sharma, Yizhou Zhang, Emilio Ferrara, and Yan Liu.
\newblock Identifying coordinated accounts on social media through hidden
  influence and group behaviours.
\newblock In {\em Proc. of ACM SIGKDD Conf. on Knowledge Discovery \& Data
  Mining}, pages 1441--1451, 2021.

\bibitem{TFIDF}
Karen Sparck~Jones.
\newblock A statistical interpretation of term specificity and its application
  in retrieval.
\newblock {\em Journal of Documentation}, 28(1):11--21, 1972.

\bibitem{tasnim2020impact}
Samia Tasnim, Md~Mahbub Hossain, and Hoimonty Mazumder.
\newblock Impact of rumors and misinformation on covid-19 in social media.
\newblock {\em Journal of Preventive Medicine and Public Health},
  53(3):171--174, 2020.

\bibitem{twitter_info_ops}
Twitter.
\newblock {Information Operations}.
\newblock
  \url{https://transparency.twitter.com/en/reports/information-operations.html},
  2022.
\newblock Accessed: 2022-06-15.

\bibitem{twitter_info_ops_2020}
{Twitter Safety}.
\newblock {Disclosing networks to our state-linked information operations
  archive}.
\newblock \url{https://t.co/etMqAUGwo2}, 2020.
\newblock Accessed: 2022-10-01.

\bibitem{twitter_info_ops_2021}
{Twitter Safety}.
\newblock {Disclosing state-linked information operations we've removed}.
\newblock
  \url{//blog.twitter.com/en_us/topics/company/2021/disclosing-state-linked-information-operations-we-ve-removed},
  2021.
\newblock Accessed: 2022-06-15.

\bibitem{vargas2020detection}
Luis Vargas, Patrick Emami, and Patrick Traynor.
\newblock On the detection of disinformation campaign activity with network
  analysis.
\newblock In {\em Proc. of ACM SIGSAC Conf. on Cloud Computing Security
  Workshop}, pages 133--146, 2020.

\bibitem{varol2017online}
Onur Varol, Emilio Ferrara, Clayton~A Davis, Filippo Menczer, and Alessandro
  Flammini.
\newblock Online human-bot interactions: Detection, estimation, and
  characterization.
\newblock In {\em Proc. Intl. AAAI Conf. on Web and Social Media (ICWSM)},
  2017.

\bibitem{wood2017does}
Zach Wood-Doughty, Michael Smith, David Broniatowski, and Mark Dredze.
\newblock {How does Twitter user behavior vary across demographic groups?}
\newblock In {\em Proc. of Workshop on NLP and Computational Social Science
  (NLP+CSS)}, pages 83--89, 2017.

\bibitem{woolley2018computational}
Samuel~C Woolley and Philip~N Howard.
\newblock {\em Computational propaganda: political parties, politicians, and
  political manipulation on social media}.
\newblock Oxford University Press, 2018.

\bibitem{yang2019arming}
Kai-Cheng Yang, Onur Varol, Clayton~A Davis, Emilio Ferrara, Alessandro
  Flammini, and Filippo Menczer.
\newblock Arming the public with artificial intelligence to counter social
  bots.
\newblock {\em Human Behavior and Emerging Technologies}, 1(1):48--61, 2019.

\bibitem{yang2020scalable}
Kai-Cheng Yang, Onur Varol, Pik-Mai Hui, and Filippo Menczer.
\newblock Scalable and generalizable social bot detection through data
  selection.
\newblock In {\em Proc. of AAAI Conf. on Artificial Intelligence (AAAI)}, pages
  1096--1103, 2020.

\bibitem{yardi2010detecting}
Sarita Yardi, Daniel Romero, Grant Schoenebeck, et~al.
\newblock {Detecting spam in a Twitter network}.
\newblock {\em First Monday}, 15(1), 2010.

\bibitem{yu2011dna}
Chenglong Yu, Mo~Deng, and Stephen S-T Yau.
\newblock Dna sequence comparison by a novel probabilistic method.
\newblock {\em Information Sciences}, 181(8):1484--1492, 2011.

\end{thebibliography}

\end{document}